\documentclass{article}
\usepackage{arxiv}
\usepackage{amsthm} 
\usepackage{amssymb}
\usepackage{amsmath}
\usepackage{empheq}
\usepackage{amscd}
\usepackage{graphicx}
\usepackage{graphics}
\usepackage[noadjust]{cite}
\usepackage{amsthm}
\usepackage{caption}
\usepackage{multirow}
\usepackage{array}
\usepackage{rotating}
\usepackage[norelsize,ruled]{algorithm2e}
\usepackage{hyperref}

\usepackage{indentfirst}
\usepackage{tikz}
\usepackage{calc}
\usepackage{epsfig}
\usepackage[numbers]{natbib}
\usepackage[makeroom]{cancel}
\usepackage{multicol}
\usepackage{csquotes}
\usepackage{enumitem}
\usepackage{nicefrac}

\usepackage{hyperref}       % hyperlinks
\hypersetup{colorlinks,linkcolor={green},citecolor={green},urlcolor={black}}
\usepackage{optidef}
\newcommand{\bolds}[1]{{#1}}   % \newcommand{\bolds}[1]{\boldsymbol{#1}}  
\renewcommand{\bold}[1]{{#1}}

\newcommand{\dip}{\psi} % rotation about y axis
\newcommand{\strike}{\phi}    % rotation about z axis
\newcommand{\tilt}{\theta}   % rotation about x axis

\newcommand{\C}[1]{\bold{C}_{#1}}  
\renewcommand{\S}[1]{\bold{S}_{#1}} 

\newcommand{\eref}[1]{equation \ref{#1}}                  % equation ref
\renewcommand{\eqref}[1]{equation \ref{#1}}                  % equation ref

\DeclareMathOperator*{\argmin}{argmin}

\begin{document}
\title{Robust Estimation of Structural Orientation Parameters and 2D/3D Local Anisotropic Tikhonov Regularization}

\author{\href{https://orcid.org/0000-0002-9879-2944}{\hspace{1mm}Ali Gholami} \\
  Institute of Geophysics, Polish Academy of Sciences, Warsaw, Poland\\
  \texttt{agholami@igf.edu.pl} \\ 
  \And 
  \href{https://orcid.org/0000-0002-9879-2944}{\hspace{1mm}Silvia Gazzola}\\
   Department of Mathematical Sciences, University of Bath, Bath, UK\\
  \texttt{S.Gazzola@bath.ac.uk} \\
  }

\graphicspath{{./figs/}}

\renewcommand{\shorttitle}{Local Anisotropic Regularization ~~~~~~~~~~~~~~~~~~ Gholami and Gazzola}

\maketitle
%%%%%%%%%%%%%%%%%%%%%%%%%%%%%%%%%%%%%%%%%%

%%%%%%%%%%%%%%%%%%%%%%%%%%%%%%%%%%%%%%%%%%
%\newpage
\begin{abstract}
Understanding the orientation of geological structures is crucial for analyzing the complexity of the Earths' subsurface. For instance, information about geological structure orientation can be incorporated into local anisotropic regularization methods as a valuable tool to stabilize the solution of inverse problems and produce geologically plausible solutions. We introduce a new variational method that employs the alternating direction method of multipliers within an alternating minimization scheme to jointly estimate orientation and model parameters in both 2D and 3D inverse problems. Specifically, the proposed approach adaptively integrates recovered
information about structural orientation, enhancing the effectiveness of anisotropic Tikhonov
regularization in recovering geophysical parameters. The paper also discusses the automatic tuning of algorithmic parameters to maximize the new method's performance. 
The proposed algorithm is tested across diverse 2D and 3D examples, including structure-oriented denoising and trace interpolation. The results show that the algorithm is robust in solving the considered large and challenging problems, alongside efficiently estimating the associated tilt field in 2D cases and the dip, strike, and tilt fields in 3D cases. Synthetic and field examples show that the proposed anisotropic regularization method produces a model with enhanced resolution and provides a more accurate representation of the true structures.
\end{abstract}
\section{Introduction}
\noindent Determining the orientation of geological structures is critical for estimating the complex characteristics of the Earth's subsurface. These key orientation parameters, including attributes like dip, strike, and tilt angles, are instrumental in revealing the nature of both linear and planar geological formations hidden beneath the Earth's surface. Such insights form the foundation for a wide range of geoscientific applications, including data processing, imaging, and interpretation. These applications perform critical tasks such as velocity analysis, velocity-independent time-domain seismic imaging, spatial data interpolation, deblending of simultaneous source data, and fault detection \citep{Li_2000_IGD,Clapp_2004_IGI,Fomel_2006_RSI,Hale_2009_SOS,Schleicher_2009_OTE,Wu_2017_DST,
Merzlikin_2020_LDI,Chen_2020_DSS,Wang_2022_MCP}.

Advanced technologies, including sophisticated data processing techniques, are used to reveal the complex nature of underground formations. Many of these processes manifest as discrete ill-posed inverse problems, as extensively documented in publications \citep{Hansen_1998_RDD,Aster_2004_PEI, Tarantola_2005_IPT}. When solving the inverse problem, anisotropic regularization is pivotal in computing solutions that integrate observed data with physical models while remaining geologically reasonable. Namely, such methods stabilize the solution by imposing regularization that depends on (an estimate of) the (local) orientation of the solution. If correct (prior) information about the orientation of structures contained in the model to be estimated is included in the regularizer, then oriented structures such as faults and steeply sloping features are suitably recovered, preventing coherent events from smearing normal to the structures.

Denoising is an important step in many data processing, inversion, and interpretation tasks.
It is used to increase the desired information content of the input signal by rejecting the unwanted part, namely noise. 
%In \citep{Bakker_1999_EPO,Hale_2009_SOS} a nonlinear structure-oriented filter was developed specifically for this purpose, via a two-step process: the structure tensors are first computed and then an anisotropic filter is designed to derive an anisotropic smoother.
A nonlinear structure-oriented filter was developed for this purpose in studies conducted by \citet{Bakker_1999_EPO} and \citet{Hale_2009_SOS}. This filter follows a two-step process: initially, structure tensors are computed, and then an anisotropic filter is designed to apply smoothing in an anisotropic manner.
An important component of anisotropic smoothers is the availability of an accurate local structural orientation field used to determine the correct direction for smoothing \citep[see, e.g.,][]{Li_2000_IGD,Clapp_2004_IGI,Fomel_2006_RSI,Hale_2009_SOS, Wang_2022_MCP}.

In two-dimensional (2D) problems, local tilt information can be extracted directly from the data, e.g., by filtering schemes \citep{Bakker_1999_EPO,Marfurt_2006_RER}. Based on the plane wave destruction algorithm \citep{Claerbout_1992_ESA}, \citet{Fomel_2002_AOP} has proposed an algorithm for estimating local event tilts/slopes directly from the data. In this approach, the tilt field appears as coefficients of plane wave prediction filters formulated as a nonlinear optimization problem in the z-transform domain and solved by the Gauss-Newton method. Fomel's algorithm has been studied and improved by a number of researchers \citep{Schleicher_2009_OTE,Chen_2013_OPW,Chen_2013_APW}. 
A major task in performing the algorithm is to stabilize the filter to mitigate sensitivity to noise in the input data \citep{Chen_2022_NLS}. To address this issue, a smoothing filter may be used in a preprocessing step to clean the input data to a certain degree before using it for the subsequent tilt estimation \citep{Wang_2022_MCP}.

Traditional filter-based algorithms for local tilt estimation can be effective for calculating the tilt field from direct measurements. However, they encounter challenges when dealing with indirect measurements where the data space does not match the model space \citep{Clapp_1998_RVE}.
In such cases, well-log information or preprocessed data that are mapped into the model space, such as stacked or migrated sections, are used in \citep{Clapp_2004_IGI} to compute the local tilt field. Subsequently, anisotropic Tikhonov regularization is applied
to solve tomographic problems.  

If information about the local tilt is available, a rotation matrix can be applied to the gradient components (horizontal and vertical derivatives) of the model so that the directional derivative along the local tilt is obtained. {The application of Tikhonov regularization with a regularization term defined with respect to the rotated gradient evaluated in a specific norm} favors solutions that have elongated features along the specified tilt angle  \citep{Li_2000_IGD,Wu_2017_SSF}. The plane-wave construction method may be used to force such features based on a model reparameterization \citep{Fomel_2006_RSI}.

In this paper, we develop a variational framework, i.e., an optimization problem, for joinlty estimating the model parameters (sometimes also referred to as the ``signal'') and the corresponding local orientation field. The basic assumption underlying the proposed method is that {a weighted norm} of the directional derivative of a signal is minimal when computed along the true local signal orientation. Therefore, in 2D space, given a clean signal, the local tilt field can be determined by minimizing {a weighted norm} of the directional derivative of the signal with respect to the local tilt field, defined between $-\pi/2$ and $\pi/2$. However, the solution is not unique in constant regions. By assuming that the signal orientation is locally stationary, we add a regularization term to the objective function to enforce the smooth variation of the tilt field. This leads to a box-constrained nonlinear minimization problem, which is efficiently solved by the alternating direction method of multipliers (ADMM) \citep{Tapia_1977_DMM,Gabay_1976_ADA,Boyd_2011_DOS}.
If the input signal is noisy or the data space does not match the model space, we solve the problem for the clean signal and the associated tilt field simultaneously using an alternating minimization approach. Specifically, given an initial tilt estimate, the clean signal is recovered through local anisotropic Tikhonov regularization defined with respect to the current tilt field estimate. The Tikhonov regularization parameter, which balances the contribution of the fit-to-data term and the reguarization term, is chosen according to the discrepancy principle, so that the restored signal fits the data up to some tolerance given by the norm of the noise in the data. 
In this way, we look for a clean signal that is close to the noisy signal but, thanks to the regularization term, has minimal fluctuations along the specified, smoothly varying local tilts. The tilt field is then updated as described above, taking as the clean signal the current estimate thereof. This process of alternating between tilt field and signal updates is repeated until a stopping criterion is satisfied.

Specifying orientation in three-dimensional (3D) space is more difficult than in 2D space, where orientation can be defined by a single tilt parameter.
We are interested in determining both linear and planar features in 3D models that can be completely specified by three angles: dip, strike, and tilt. A {weighted norm} of the directional derivatives in two orthogonal directions is minimized to determine these angles.
{Similarly to the 2D case,} we propose 3D regularization functions that are structure adaptive and can properly regularize 3D inverse problems based on this information. The ADMM is used to minimize the objective function while simultaneously estimating the model and orientation parameters.

Although in this paper we are interested only in the regularized output of our 2D and 3D algorithms, the estimated orientation parameters also can be used as a side-product for other purposes. For example, the orientation parameters have important seismic applications for interpretation and imaging (see, e.g., \citet{Marfurt_2006_RER,Fomel_2007_VIT,Schleicher_2009_OTE,Wu_2017_DST,Merzlikin_2020_LDI}).

The main contributions of this work can be summarized as follows:
\begin{enumerate}
\item We introduce a novel optimization problem designed for the estimation of local orientation fields, including tilt in 2D signals and dip, strike, and tilt in 3D signals. Additionally, we present an efficient computational algorithm adapted to solve this optimization problem.
\item We propose an algorithm that, through an alternating minimization strategy, naturally incorporates the newly developed local orientation estimation scheme (taylored to address 2D and 3D geophysical inverse problems) into the framework of adaptive anisotropic regularization.
\item We apply the formulated algorithms to geophysical applications, specifically focusing on denoising and trace interpolation. The results obtained from these applications demonstrate the practical utility and effectiveness of the algorithm in tackling large-scale and challenging data processing problems.
\end{enumerate}

The rest of the paper is organized as follows. First, assuming local tilt information is available, we present the local anisotropic Tikhonov regularization method for 2D problems. We then set up an optimization problem that allows for the simultaneous estimation of local tilts and model parameters, and we explain how it can be solved using alternating minimization and ADMM. Then we extend our method to 3D problems and show some experimental results for the 2D and 3D cases. We end with some concluding remarks.

\section{2D Anisotropic regularization}
% When information regarding the structural orientation of an expected model becomes accessible, it can serve as valuable a priori knowledge, enhancing the effectiveness of anisotropic regularization techniques in generating geologically plausible models \citep{Li_2000_IGD,Clapp_2004_IGI}. Yet, securing such insights can pose formidable challenges, especially in cases involving indirect measurements. Consequently, the estimation of structural orientation parameters and the investigation of the unknown model become intertwined, necessitating a simultaneous determination.

\begin{figure}
\center
\includegraphics[width=1\columnwidth]{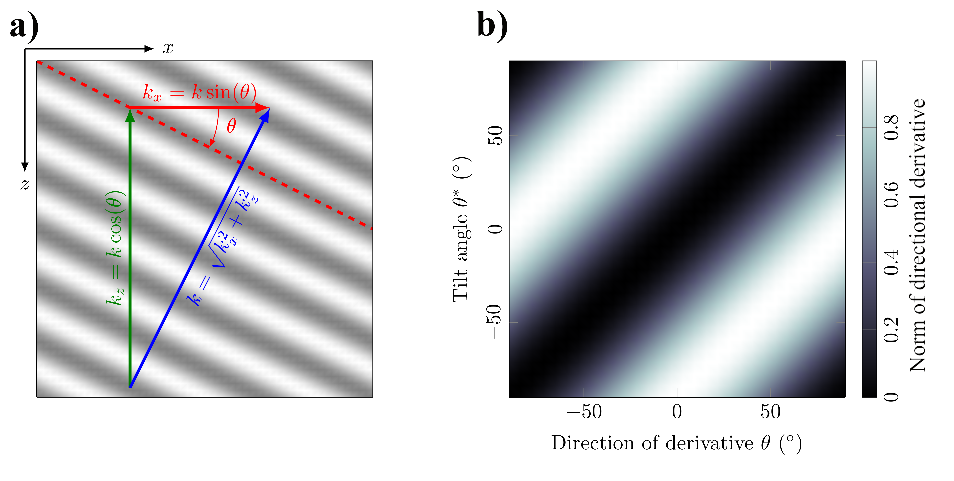}
\caption{(a) A 2D sinusoidal plane wave $m(x,z)= \cos(k_xx-k_zz):=\cos(k \sin(\theta) x-k\cos(\theta)z)$ described by the tilt angle pointing in the direction of zero data gradient (constant phase), perpendicular to the wavenumber $k$ that is in the direction of highest data gradient (maximum phase). 
(b) Norm of directional derivative $\|\nabla_{\!\tilt}\cos(k \sin(\theta^*) x-k\cos(\theta^*)z)\|_2^2$ as a function of the signal tilt and the direction of derivative.
}
\label{slope2}
\end{figure}

% In this paper we consider both 2D and 3D problems; first the 2D case is considered and then the algorithm is extended to the 3D case. 
%
In the 2D case, the model parameters vector $\bold{m}$, representing a quantity of interest, is obtained by stacking the columns of a rectangular $N_z\times N_x$ array such that its length $N$ is given by the product $N_zN_x$.
In various steps of data processing and imaging, such models may appear as the solution of ill-conditioned linear systems of equations of the form 
\begin{equation}\label{main_eq}
\bold{d=Gm + e},
\end{equation}
where $\bold{d}$ and $\bold{e}$ are real vectors of length $M$ representing known measured data and unknown white Gaussian noise, respectively, the unknown $\bold{m}$ is a real vector of length $N$, and
$\bold{G}$ is the {linear} forward operator from the model space to the data space \citep{Hansen_1998_RDD,Aster_2004_PEI, Tarantola_2005_IPT}. Due to the presence of noise and the typical ill-conditioning of the system, a suitable regularization is required to produce a reasonable estimate of the model. Tikhonov regularization defines $\bold{m}$ as the solution of the following optimization problem:
\begin{equation}\label{eq:iso}
 \min_{\bold{m}} \frac{\mu}{2}\|\bold{G}\bold{m}-\bold{d}\|_2^2+ \frac12\sum_{i=1}^N\|(\nabla \bold{m})_i\|_2^2,
\end{equation}
where the penalty parameter $\mu$ balances the action of the data fit against regularization, $\|\cdot\|_2$ denotes the vector $2$-norm, and  $\nabla$ is a scaled discrete gradient operator.
We recall that $(\nabla\bold{m})_i$ is a two-component vector of the form
\begin{equation}
(\nabla\bold{m})_i=
\begin{bmatrix}
(\nabla_{\!x}\bold{m})_i\\
(\nabla_{\!z}\bold{m})_i
\end{bmatrix},
\end{equation}
where $\nabla_{\!x}$ and $\nabla_{\!z}$ are finite difference operators discretizing the first-order horizontal ($x$) and vertical ($z$) derivatives.
A main issue of the Tikhonov regularizer is that it is homogeneous, in the sense that it does not take into account the structure of the regularized solution and therefore tends to over-smooth the model. One possible way to overcome this problem is to implement anisotropic regularization by penalizing directional derivatives. This can be done by rotating and scaling the gradient vector in the regularization function as follows:
\begin{equation} \label{AnisoReg}
\sum_{i=1}^N\| \bold{R}(\bolds{\tilt}_i)(\nabla \bold{m})_i\|_{\bold{\Sigma}_{{i}}}^2,
\end{equation} 
where
\begin{equation}
\bold{R}(\bolds{\tilt}_i)=
\begin{pmatrix}
~~\cos\bolds{\tilt}_i & \sin\bolds{\tilt}_i\\
-\sin\bolds{\tilt}_i & \cos\bolds{\tilt}_i
\end{pmatrix}:=
\begin{pmatrix}
\bold{x}'(\bolds{\tilt}_i)\\
\bold{z}'(\bolds{\tilt}_i)
\end{pmatrix},
\end{equation}
and 
$
\bolds{\Sigma}_{{i}}=
\begin{pmatrix}
\sigma_{x} & 0\\
0 & \sigma_{z}
\end{pmatrix}$.
Here and below, $\bolds{\tilt}_i \in (-\frac{\pi}{2},\frac{\pi}{2}]$ is the local signal tilt angle (orientation) which is measured clockwise about the origin (see Figure \ref{slope2}), $\bold{R}$ is a rotation matrix that converts any coordinates in the original ($x,z$) system into ($x',z'$) coordinates defined in terms of $\bolds{\tilt}_i$, and $\|\bold{g}\|_{\bolds{\Sigma}_{{i}}}^2=\bold{g}^T{\bolds{\Sigma}_{{i}}}\bold{g}$ denotes the weighted vector $2$-norm with weighting matrix $\bolds{\Sigma}_{{i}}$ that weights the rotated gradient components by the coefficients  $\sigma_{x}\geq \sigma_{z} \geq 0$.
This weighting allows for flexible smoothing of the signal in the tilt direction relative to the orthogonal one, as well as control over the anisotropic behavior of the regularization.
For $\sigma_{z}=0$, no smoothing orthogonal to the $\bolds{\tilt}_i$ direction is applied. The regularization in this case is maximally anisotropic and favors models with elongated features along the structure direction $\theta_i$.
In general, $\sigma_{x}$ and $\sigma_{z}$ can be adjusted locally for each model point \citep{Lavialle_2007_SFP}. However, for the sake of simplicity, we will use fixed values in this paper {(so that $\bolds{\Sigma}_{{i}}=\bolds{\Sigma}$)}. In this case, without loss of generality we set $\sigma_{x}=1$ and $\sigma_{z}=\epsilon$ with $\epsilon \ll 1$, forcing line-like linear structures in the final model.

As stated in the Introduction, our approach to estimating the tilt field hinges on a fundamental hypothesis: an estimate of the tilt field for a 2D model can be obtained by minimizing a weighted norm of the directional derivative defined with respect to such tilt field. In Appendix A, we show that this is true for a simple 2D sinusoidal plane wave.

In general, given an initial guess $\tilt$ for the tilt field, we propose an iterative method that simultaneously estimates both the tilt field and model parameters by the following two-step alternating minimization scheme
\begin{align} 
&\bold{m}^{+}=\argmin_{\bold{m}} \frac{\mu}{2}\|\bold{G}\bold{m}-\bold{d}\|_2^2+ \frac12\!\sum_{i=1}^N\| \bold{R}(\bolds{\tilt}_i)(\nabla \bold{m})_i\|_{\bold{\Sigma}}^2,\label{main_m}\\
&\bolds{\tilt}^{+}=\argmin_{\bolds{\tilt}\in B} \frac12\sum_{i=1}^N\| \bold{R}(\bolds{\tilt}_i)(\tilde{\nabla}\bold{m}^{+})_i\|_{\bold{\Sigma}}^2+ \frac{\rho}{2}\|\nabla\bolds{\tilt}\|_2^2, \label{main_theta}
\end{align}
where $\bolds{\tilt}^{+}$ is the updated tilt associated with the updated model parameters $\bold{m}^{+}$. Specifically, at each iteration of the above scheme, the first subproblem in \eqref{main_m} involves the estimation of the model parameters by local anisotropic Tikhonov regularization, given the most recent tilt field $\bolds{\tilt}$, while the second subproblem in \eqref{main_theta} involves the update of the tilt field based on the most recently estimated model $\bold{m}^+$. Here,  $B=\{\bolds{\tilt}\in \mathbb{R}^N: -\frac{\pi}{2}<\bolds{\tilt} \leq\frac{\pi}{2}\}$ and $\rho>0$ is a positive parameter controlling the smoothness of the estimated tilt field.
In \eqref{main_theta}, $\tilde{\nabla}\bold{m}$ represents a smoothed (regularized) estimate of the model gradient $\nabla \bold{m}$. We achieve this by smoothing the gradient components using smooth derivative filters $\bold{h}_x$ and $\bold{h}_z$, i.e.
\begin{equation} \label{dxz}
\tilde{\bold{m}}_x:=\tilde{\nabla}_x\bold{m}=\bold{h}_x\star \bold{m},\quad \tilde{\bold{m}}_z:=\tilde{\nabla}_z\bold{m}=\bold{h}_z \star\bold{m},
\end{equation} 
where $\star$ denotes the 2D convolution operator. These filters, for the conventional first-order finite difference approximation (without smoothing), are defined as:
\begin{equation} \label{diffs}
\bold{h}_x=
\begin{bmatrix}
\frac{1}{2} & 0& -\frac{1}{2}
\end{bmatrix},
\quad
\bold{h}_z=
\begin{bmatrix}
\frac{1}{2} &
0&
 -\frac{1}{2}
\end{bmatrix}^T.
\end{equation}
While these filters provide a critical $90^{\circ}$ phase shift that facilitates orientation determination, they also introduce unwanted amplitude scaling issues. To mitigate these instabilities and obtain a stable estimate of signal variations, we remove the frequency-dependent amplitude scaling from the derivatives. 
For this purpose, we employ the Hilbert transform, known for its $90^{\circ}$ phase shift property akin to derivatives but without altering amplitudes \citep{Claerbout_1976_FGD}. The associated filters are defined as \citep{Felsberg_2000_NEL}:
\begin{equation}
h_x = -\frac{1}{2\pi}\frac{x}{(x^2+z^2)^{\nicefrac{3}{2}}},\quad
h_z = -\frac{1}{2\pi}\frac{z}{(x^2+z^2)^{\nicefrac{3}{2}}}.
\end{equation}
These filters are illustrated in Figure \ref{Riesz_filters}. They represent a smoothed version of the differentiators in \eqref{diffs}.
To show the performance of these filters, we present a comparison in Figure \ref{Riesz_trans}. A sinusoidal plane wave both with and without added random noise is shown in Figure \ref{Riesz_trans}a and b, respectively. We computed the horizontal derivative of the signal obtained using the Hilbert transform filter (Figure \ref{Riesz_trans}c and d) and the standard finite differences method (Figure \ref{Riesz_trans}e and f). We observe that, the Hilbert transform filter exhibits greater stability in the presence of noise.

\subsection{Solving the Model Subproblem in \eqref{main_m}}
We first note that
\begin{equation} \label{D2d_nln}
 \sum_{i=1}^N\| \bold{R}(\bolds{\tilt}_i)(\nabla \bold{m})_i\|_{\bold{\Sigma}}^2 = \|\bold{D}(\bolds{\tilt})\bold{m}\|_2^2,
\end{equation}
where $\bold{D}(\bolds{\tilt})$, with $\theta=[\theta_1,\dots,\theta_N]^T$, represents the local anisotropic differential operator, given by
\begin{equation} \label{D_aniso}
\bold{D}(\bolds{\tilt}) = 
\begin{pmatrix}
\bold{I} & 0\\
0 & \sqrt{\epsilon}\bold{I}
\end{pmatrix}
\begin{pmatrix}
\C{\bolds{\tilt}} & \S{\bolds{\tilt}}\\
-\S{\bolds{\tilt}} & \C{\bolds{\tilt}}
\end{pmatrix}
\begin{pmatrix}
\nabla_{\!x}\\
\nabla_{\!z}
\end{pmatrix}.
\end{equation}
Here $\C{\bolds{\tilt}}$ and $\S{\bolds{\tilt}}$ denote diagonal matrices with $\cos\bolds{\tilt}$ and $\sin\bolds{\tilt}$ on their main diagonal, respectively. 
Therefore, the minimization problem in \eqref{main_m} is quadratic with respect to $\bold{m}$. Applying the optimality conditions leads us to the following expression for the updated model parameters:
\begin{equation} \label{m_est}
\bold{m}^+ = \left(\mu\bold{G}^T\bold{G}+\bold{D}(\bolds{\tilt})^T\bold{D}(\bolds{\tilt})\right)^{-1}
(\mu\bold{G}^T\bold{d}).
\end{equation}

%%%%%%%%%%%%%%%%%%%%%%%%%%%%%%%%%%%%%%%%%%%%%%%%%%%%%%%%%%%%%%%%%%%%%%%%%%%%%%%%%%%%%%%%%%%%%%%%
%%%%%%%%%%%%%%%%%%%%%%%%%%%%%%%%%%%%%%%%%%%%%%%%%%%%%%%%%%%%%%%%%%%%%%%%%%%%%%%%%%%%%%%%%%%%%%%%
\subsection{Solving the Tilt Subproblem in \eqref{main_theta}}\label{sect:tilt2D}
The problem in \eqref{main_theta} is a constrained nonconvex and nonlinear optimization problem with respect to $\bolds{\tilt}$, thus iterative gradient-based methods can be employed to (locally) solve it.
Introducing the auxiliary variable $\bold{z}$, the constraint $\bolds{z}=\bolds{\tilt}$, and the Lagrange multiplier vector $\bolds{\lambda}$, the ADMM  \citep{Tapia_1977_DMM,Gabay_1976_ADA,Boyd_2011_DOS} leads to the following iteration (see Appendix B):
\begin{subequations} \label{theta}
\begin{align} 
\bolds{\tilt}^{+}&=(\bold{J}(\bold{m}^{\!+}\!\!,\bolds{\tilt})^T\!\bold{J}(\bold{m}^{\!+}\!\!,\bolds{\tilt})+\tau I+\rho \nabla^T\nabla)^{-1}\!
\bold{g}(\bold{m}^{\!+}\!\!,\bolds{\tilt},\bold{z},\bolds{\lambda}), \label{sub_theta} \\
\bold{z}^{+} &= \min(\max(\bolds{\tilt}^+-\bolds{\lambda},-\frac{\pi}{2}),\frac{\pi}{2}), \\
\bolds{\lambda}^{+}& =\bolds{\lambda} - \tau(\bolds{\tilt}^{+}-\bold{z}^{+}),
\end{align}
\end{subequations}
where {$J(m,\theta)$ denotes the Jacobian of $\bold{D}(\bolds{\tilt})\bold{m}$,  
%the term $\sum_{i=1}^N\| \bold{R}(\bolds{\tilt}_i)(\tilde{\nabla}\bold{m})_i\|_{\bold{\Sigma}}^2$}, 
the triplet $(\bolds{\tilt},\bold{z},\bolds{\lambda})$ are the values at the current iteration, $(\bolds{\tilt}^{+},\bold{z}^{+},\bolds{\lambda}^{+})$ are the updated values, and $\tau$ is the penalty parameter. 

%%%%%%%%%%%%%%%%%%%%%%%%%%%%%%%%%%%%%%%%%%%%%%%%%%%%%%%%%%%%%%%%%%%%%%%%%%%%%%%%%%%%%%%%%%%%%%%%
\begin{figure}
\center
\includegraphics[width=1\columnwidth]{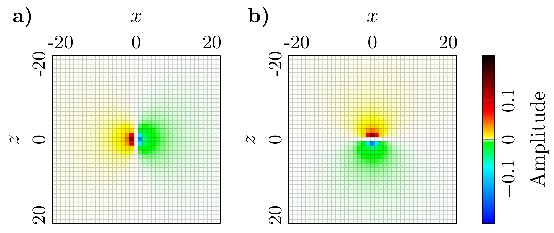}
\caption{The 2D Hilbert transform filters in (a) the horizontal direction and (b) the vertical direction.
}
\label{Riesz_filters}
\end{figure}
%%%%%%%%%%%%%%%%%%%%%%%%%%%%%%%%%%%%%%%%%%%%%%%%%%%%%%%%%%%%%%%%%%%%%%%%%%%%%%%%%%%%%%%%%%%%%%%%

%%%%%%%%%%%%%%%%%%%%%%%%%%%%%%%%%%%%%%%%%%%%%%%%%%%%%%%%%%%%%%%%%%%%%%%%%%%%%%%%%%%%%%%%%%%%%%%%
\begin{figure}
\center
\includegraphics[width=0.5\columnwidth]{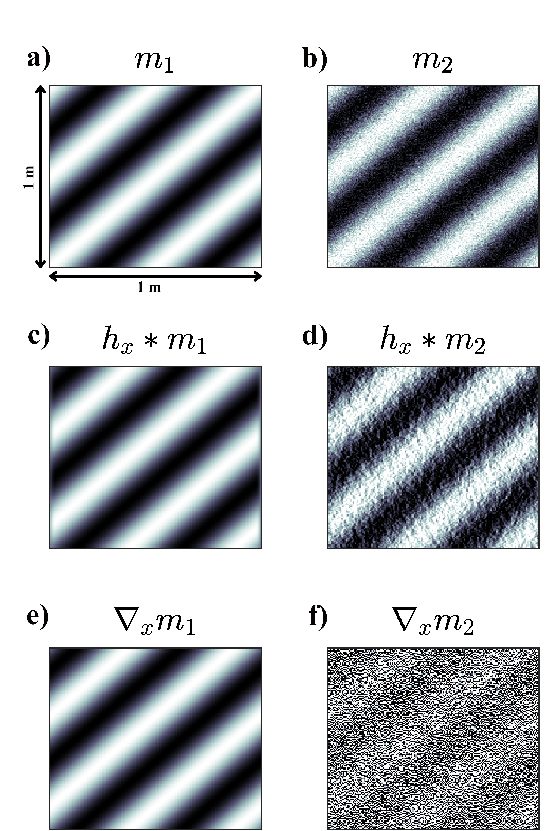}
\caption{Comparison of signal derivatives calculated using the Hilbert transform filter (c-d) and standard finite differences (e-f) applied to a sinusoidal plane wave with and without added random noise (a-b).
}
\label{Riesz_trans}
\end{figure}
%%%%%%%%%%%%%%%%%%%%%%%%%%%%%%%%%%%%%%%%%%%%%%%%%%%%%%%%%%%%%%%%%%%%%%%%%%%%%%%%%%%%%%%%%%%%%%%%

\subsection{Selection of the regularization parameters}
The parameters $\mu$, $\rho$, and $\tau$ should be properly set for the proposed algorithm to produce meaningful results. We dynamically tune the value of these parameters during the iterations.

Regarding the regularization parameter $\mu$ for anisotropic Tikhonov regularization applied to the model parameters $m$, we assume that a good estimate of the data error bound $\varepsilon$, defined as
\begin{equation} \label{circ_cons}
\|\bold{Gm}-\bold{d}\|_2^2=\varepsilon,
\end{equation}
is available. 
%Since we have separate optimization problems over $\bold{m}$ and $\bolds{\tilt}$ we may consider that $\bold{m}^{k}$ depends only on $\mu$.
This allows us to update $\mu$ by a fixed-point iteration solving for the constraint in \eqref{circ_cons}:
\begin{equation} \label{alpha_update}
\mu^{+}=\dfrac{2\|\bold{Gm}^+-\bold{d}\|_2^2}{\|\bold{Gm}^+-\bold{d}\|_2^2+\varepsilon}\mu,
\end{equation}
where $\mu^+$ is the updated parameter and $\bold{m}^+$ is the model obtained by using the current parameters $(\mu,\rho,\tau)$; see \citet{Gholami_2022_ABP}.

{Choosing an optimal value for the regularization parameter for the tilt field $\rho$ is challenging. 
% as it determines the smoothness of the tilt field in the regularization function. 
While this paper does not aim to provide a method for selecting an optimal value for $\rho$, a practical approach is offered to simplify the search. At each iteration, $\rho$ is set proportionally to the maximum value of the vector diag($\bold{J}^T\bold{J}$), ensuring reasonable smoothness of the tilt field and minimization of the first term in the anisotropic regularization term. Specifically
\begin{equation} \label{rho}
\rho = \epsilon_1\max(\text{diag}(\bold{J}^T\bold{J})),
\end{equation}
where $\epsilon_1$ is a positive parameter.

The parameter $\tau$ is associated to the ADMM method used to solve the optimization problem defined in \eref{main_theta}, and serves as a penalty parameter enforcing the tilt field to lie between $-\pi/2$ and $\pi/2$. At each iteration, its value is set proportional to the maximum value of the vector diag($\bold{J}^T\bold{J}$), achieving a reasonable damping towards the box constraint and minimizing the first term in the anisotropic regularization term. Specifically:
\begin{equation} \label{tau}
\tau= \epsilon_2\max(\text{diag}(\bold{J}^T\bold{J})),
\end{equation}
where $\epsilon_2$ is a positive parameter.

For the 2D numerical examples in this paper, we use $\epsilon_1=1$ and $\epsilon_2=0.1$.
%We provide values for $\epsilon_1$ and $\epsilon_2$ in Section ``Numerical Examples" that often work well in practice.

The proposed algorithm for 2D problems is summarized in Algorithm \ref{Algorithm2D}. 
% Note that, referring to the notations in Algorithm \ref{Algorithm2D}, the approach described in \eref{theta} sets $K=1$. 
}
%-----------------------------------------------------------------------------------
%--------------------------------------------- 2D algorithm
%-----------------------------------------------------------------------------------
\begin{algorithm}[!h]
\vspace{.2cm}
 \caption{Local anisotropic regularization with tilt estimation in 2D space.}  \label{Algorithm2D}
 Inputs: data $\bold{d}$, forward operator $\bold{G}$, initial penalty parameter $\mu$, $\epsilon_1$, and $\epsilon_2$.\\ 
 Set  $\bolds{\tilt}=\bold{z}=\bolds{\lambda}=\bold{0}$.   \\
 \While{\text{a stopping criterion is {not} satisfied} }
 {
 %--------------------------------------------------------------------------------------
%\% update the model\\
 $\bold{m}\leftarrow \left(\mu\bold{G}^T\bold{G}+\bold{D}(\bolds{\tilt})^T\bold{D}(\bolds{\tilt})\right)^{-1}(\mu\bold{G}^T\bold{d})$ \\
 \vspace{2mm} 
%-------------------------------------------------------------------------------------
\For{\text{$k=1,...,K$}}{
 \vspace{2mm}
%--------------------------------------------------------------------------------------
%\% update the tilt angle \\
$\bolds{\tilt}\leftarrow \!(\bold{J}(\bold{m},\!\bolds{\tilt})^T\!\!\bold{J}(\bold{m},\!\bolds{\tilt})+\tau I+\!\rho \nabla^T\nabla)^{-1}
\bold{g}(\bold{m},\!\bolds{\tilt},\bold{z},\bolds{\lambda}) $\\  \vspace{2mm} 
%--------------------------------------------------------------------------------------
%\% update the auxiliary variable \\
$\bold{z}\leftarrow \min(\max(\bolds{\tilt}-\bolds{\lambda},-\frac{\pi}{2}),\frac{\pi}{2})$ \\ \vspace{2mm} 
%--------------------------------------------------------------------------------------
%\% update the Lagrange multiplier \\
$\bolds{\lambda}\leftarrow\bolds{\lambda}- \tau(\bolds{\tilt}-\bold{z}$)\\ \vspace{2mm}
}
%--------------------------------------------------------------------------------------
%\% update the data fit parameter \\
$\mu \leftarrow \dfrac{2\|\bold{Gm}-\bold{d}\|_2^2}{\|\bold{Gm}-\bold{d}\|_2^2+\varepsilon}\mu$
}
Outputs: model $\bold{m}$ and tilt field $\bolds{\tilt}$
\end{algorithm}

%%%%%%%%%%%%%%%%%%%%%%%%%%%%%%%%%%%%%%%%%%%%%%%%%%%%%%%%%%%%%%%%%%%%%%%%%%%%%%%%%%%%%%%%%%%%%%%%
\begin{figure}
\center
\includegraphics[width=0.8\columnwidth]{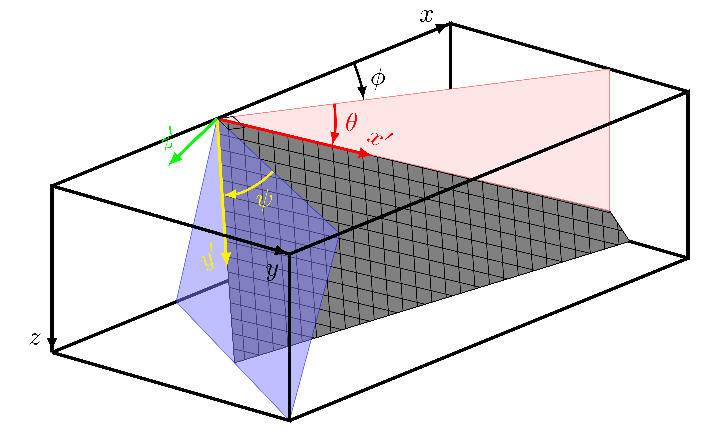}
\caption{Quantities from mathematics and geology used to specify the orientation of linear and planar events in 3D space. A linear event (shown by the vector $\bold{x}'$) is mathematically determined by the two angles strike ($\strike$) and tilt ($\tilt$). The tilt ($\tilt$) represents the angle, measured from the horizon, of the direction of maximum coherency in the vertical plane (in red) of the strike angle ($\strike$), which is measured from the positive $x$ axis. The dip ($\dip$) is the angle measured from the horizon of the maximum coherency (shown by the vector $\bold{y}'$) in a plane (in blue) perpendicular to $\bold{x}'$. 
Planar events as formed by two vectors $\bold{x}'$ and $\bold{y}'$ and are determined by three angles ($\strike,\tilt,\dip$). 
}
\label{3D_space}
\end{figure}
%%%%%%%%%%%%%%%%%%%%%%%%%%%%%%%%%%%%%%%%%%%%%%%%%%%%%%%%%%%%%%%%%%%%%%%%%%%%%%%%%%%%%%%%%%%%%%%%
%\section{formulations in 3D space}

\section{3D anisotropic regularization}
Navigating orientation in 3D space is naturally more difficult than in 2D, where a single parameter, the tilt $\tilt$, may simply specify the orientation of linear events (see Figure \ref{slope2}). To deal with this complication, we use a right-handed coordinate system, with the $x$ axis extending forward, the $y$ axis pointing rightward, and the $z$ axis downward (see Figure \ref{3D_space}).

In 3D space, the directional derivative of the model $\bold{m}$ in the direction of a vector
\begin{equation}
\bold{x}'(\alpha,\beta,\gamma)=[\cos\alpha,\cos\beta,\cos\gamma],
\end{equation}
takes the form:
\begin{equation} \label{dir_deriv_3D}
\nabla_{\bold{x}'}\bold{m} = \bold{m}_x\cos\alpha + \bold{m}_y\cos\beta + \bold{m}_z\cos\gamma 
=\bold{x}'(\alpha,\beta,\gamma)\cdot \nabla\bold{m},
\end{equation}
where $\alpha, \beta$, and $\gamma$ denote the direction angles formed by the vector and the positive $x$, $y$, and $z$ axes, respectively, $\cdot$ denotes the Euclidean inner product in $\mathbb{R}^3$, and $\nabla\bold{m}=[m_x,m_y,m_z]^T$. These angles offer a comprehensive means of characterizing the direction of vectors in 3D space.
We simplify our notation to align with geology and geophysics conventions by using two independent angles, the strike ($\strike$) and tilt ($\tilt$), derived from the spherical coordinate system. These angles relate to the direction angles as follows (using the relation between direction cosines):
\begin{equation} \label{dip_az}
\begin{cases}
\cos\alpha=\cos\tilt\cos\strike,\\
\cos\beta=\cos\tilt\sin\strike, \\
\cos\gamma=\sin\tilt.
\end{cases}
\end{equation}
{  
These relations facilitate a more consistent notation between our 2D and 3D formulations and within the context of geology and geophysics, although alternative relations are also possible, leading to different conventions. 
}

Figure \ref{3D_space} shows a tilted vector $\bold{x}'$, representing a linear event in a dipping plane contained in the model, and the angles $\strike$ and $\tilt$. 
The strike angle $\strike$, measured from the positive $x$ axis, specifies the direction of the vertical plane that includes the tilt vector and 
$\tilt$ is the inclination angle of the vector from the horizon. 
By using these new angles we rewrite the directional derivative in \eqref{dir_deriv_3D} as
\begin{equation*} \label{direc_deriv3d}
\nabla_{\!\bold{x}'\!}\bold{m} =\bold{m}_x\cos\tilt\cos\strike +\bold{m}_y\cos\tilt\sin\strike + \bold{m}_z\sin\tilt .
\end{equation*}
Similar to the 2D case we can determine the local dip and local strike angles at each point of a 3D model by minimizing 
$\|\nabla_{\bold{x}'}\bold{m}\|_2^2$. 
%However, before diving into the optimization procedure, we provide some mathematical and geological analysis of these angles. 
However, we can only use this equation to find the direction of greatest coherence in a sphere centered at each point.
This only determines linear events in 3D space and thus cannot determine the orientation of the 2D planar features. 
To do this we need also to specify linear events in the direction perpendicular to the direction
\begin{equation}
\bold{x}'(\strike,\tilt)=
(\cos\tilt\cos\strike, \cos\tilt\sin\strike, \sin\tilt).
\end{equation}
This necessitates calculating the derivative orthogonal to $\bold{x}'$ and searching for the maximum coherence in the plane perpendicular to this direction (the blue plane in Figure \ref{3D_space}).
As a result, the first step is to specify the plane perpendicular to $\bold{x}'$. This can be accomplished in a variety of ways. The most basic one is to use the Euler angle system.
The vector $\bold{x}'$ in Figure \ref{3D_space} can be expressed in a rotated coordinate system created by rotating the original coordinate system about the $z$ axis by $\strike$ and then rotating the outcomes about the $y$ axis by $\tilt$. These rotations can be performed using the $3\times 3$ rotation matrices $\bold{R}_y(\tilt)$ and $\bold{R}_z(\strike)$ defined as
\begin{equation}
\bold{R}_y(\tilt)=
\begin{pmatrix}
\cos\tilt & 0 & \sin\tilt \\
 0 & 1 & 0\\
 -\sin\tilt & 0 & \cos\tilt 
\end{pmatrix},
\end{equation}
and
\begin{equation}
\bold{R}_z(\strike)=
\begin{pmatrix}
\cos\strike & \sin\strike & 0\\
-\sin\strike & \cos\strike & 0\\
 0 & 0 & 1
\end{pmatrix},
\end{equation}
giving the total rotation matrix 
\begin{equation*}
\bold{R}_y(\tilt)\bold{R}_z(\strike)\!=\!
\begin{pmatrix}
\cos\tilt\cos\strike & \cos\tilt\sin\strike& \sin\tilt \\
-\sin\strike & \cos\strike &  0\\
-\sin\tilt\sin\strike  & -\sin\tilt\cos\strike  & \cos\tilt\\
\end{pmatrix}.
\end{equation*}
As can be seen, the first row of this matrix provides the desired direction $\bold{x}'$,  and because the rotation matrices are orthonormal, the second and third rows of this matrix form the basis for the desired plane perpendicular to $\bold{x}'$.
This plane can be rotated using the rotation matrix
\begin{equation}
\bold{R}_x(\dip)=
\begin{pmatrix}
 1 & 0 & 0\\
0 &\cos\dip & \sin\dip \\
0 & -\sin\dip & \cos\dip
\end{pmatrix}.
\end{equation}
The rotation angle $\dip$ is clockwise about the $\bold{x}'$ axis. 
% by $\dip$, when viewed along it. 
The total rotation matrix is defined by
{\small
\begin{align}  \label{R3D}
\bold{R}(\strike,\tilt,\dip)
&=\bold{R}_x(\dip)\bold{R}_y(\tilt)\bold{R}_z(\strike):=
\begin{pmatrix}
\bold{x}'(\strike,\tilt,\dip)\\
\bold{y}'(\strike,\tilt,\dip)\\
\bold{z}'(\strike,\tilt,\dip)
\end{pmatrix}\nonumber \\
&=\begin{pmatrix}
\cos\tilt\cos\strike  & \cos\tilt\sin\strike  & \sin\tilt\\
%---------------------------------------------------------------------------------------
-\cos\dip \sin\strike- \sin\dip\sin\tilt\cos\strike & 
\cos\dip\cos\strike -\sin\dip\sin\tilt\sin\strike & \sin\dip\cos\tilt\\
%---------------------------------------------------------------------------------------
\sin\dip\sin\strike - \cos\dip\sin\tilt\cos\strike &
-\cos\dip\sin\strike - \cos\dip\sin\tilt\sin\strike& \cos\dip\cos\tilt
\end{pmatrix}.
\end{align}
}
%%%%%%%%%%%%%%%%%%%%%%%%%%%%%%%%%%%%%%%%%%%%%%%%%%%%%%%%
%\begin{equation}
%\begin{aligned}[c] \label{R3D}
%&\bold{R}(\strike,\tilt,\dip)=\bold{R}_x(\dip)\bold{R}_y(\tilt)\bold{R}_z(\strike)
%&\begin{pmatrix}
%\cos(\tilt)\cos(\strike)  & \cos(\tilt )\sin(\strike )  & \sin(\tilt)\\
%---------------------------------------------------------------------------------------
%-\cos(\dip) \sin(\strike)- \sin(\dip)\sin(\tilt)\cos(\strike) & 
%\cos(\dip)\cos(\strike) -\sin(\dip)\sin(\tilt)\sin(\strike) & \sin(\dip)\cos(\tilt)\\
%---------------------------------------------------------------------------------------
%\sin(\dip)\sin(\strike) - \cos(\dip)\sin(\tilt)\cos(\strike) &
%-\cos(\dip)\sin(\strike) - \cos(\dip)\sin(\tilt)\sin(\strike)& \cos(\dip)\cos(\tilt)
%\end{pmatrix} \nonumber\\
%:=
%\begin{pmatrix}
%\bold{x}'(\strike,\tilt,\dip)\\
%\bold{y}'(\strike,\tilt,\dip)\\
%\bold{z}'(\strike,\tilt,\dip)
%\end{pmatrix}.
%\end{aligned}
%\end{equation}

{In this formulation, the rotated coordinate system ($x',y',z'$) is obtained by rotating the original coordinate system ($x,y,z$) using three successive rotations $\bold{R}_z(\strike)$, $\bold{R}_y(\tilt)$, and $\bold{R}_x(\dip)$. 
In this notation, the tilt ($\tilt$) represents the angle, measured from the horizontal, of the direction of maximum coherency in the vertical plane of the strike angle ($\strike$), which is measured from the positive $x$ axis. On the other hand, the dip ($\dip$) is the angle, measured from the horizontal, of the maximum coherency in a plane perpendicular to the direction of maximum coherency.
}

We define the local anisotropic regularization term for 3D problems using the 3D rotation matrix in \eqref{R3D} as
\begin{equation} \label{AnisoReg3D}
\sum_{i=1}^N\|\bold{R}(\bolds{\strike}_i,\bolds{\tilt}_i,\bolds{\dip}_i)(\nabla \bold{m})_i\|_{\bolds{\Sigma}_{{i}}}^2,
\end{equation} 
where 
\begin{equation}
(\nabla\bold{m})_i=
\begin{pmatrix}
(\nabla_{\!x}\bold{m})_i\\
(\nabla_{\!y}\bold{m})_i\\
(\nabla_{\!z}\bold{m})_i
\end{pmatrix}, \quad
\bolds{\Sigma}_i=
\begin{pmatrix}
\sigma_{x} & 0 & 0\\
0 & \sigma_{y} & 0\\
0 & 0 & \sigma_{z}
\end{pmatrix}.
\end{equation}
The degree of anisotropy in our regularization is controlled through the parameters $\sigma_{x} \geq \sigma_{y} \geq \sigma_{z} \geq 0$ {for every point in the domain 
(so that $\bolds{\Sigma_{{i}}}=\bolds{\Sigma}$)}. These parameters can be adjusted individually for each point in the model, as demonstrated by \citet{Lavialle_2007_SFP}. However, for the sake of simplicity and to maintain consistency, we employ fixed values throughout this paper. Without any loss of generality, we set $\sigma_{x}=1, \sigma_{y}=\delta$, and $\sigma_{z}=\epsilon$, where $1> \delta > \epsilon$, with $\epsilon \ll 1$. This specific configuration enforces a preference for planar-like and line-like structures within the model.  {In this case, the unit vector $\bold{z}'$ will indicate the directions with the highest model gradients while the unit vector $\bold{x}'$  will indicate the directions with the lowest model gradients, aligned along linear features. The unit vector $\bold{y}'$ will indicate the directions with intermediate data gradients. Planar features in the model will be explained by the $\bold{x}'\bold{y}'$ plane.} Figure \ref{ellipsoinds} depicts an anisotropy ellipsoid at a model point. The highest degree of anisotropy occurs when $\sigma_y=\sigma_z=0$, resulting in a needle-shaped ellipsoid that favors models with elongated features aligned with the $x'$ direction. In the case of isotropy, where all parameters are set to unity, the regularization behaves spherically.

{Similarly to the 2D case,} to address 3D linear inverse problems we regularize the associated linear system $\bold{Gm}=\bold{d}$ using a local anisotropic Tikhonov method. 
Given an initial guess $(\bolds{\strike},\bolds{\tilt},\bolds{\dip})$ for the orientation parameters, this method employs a two-step iterative procedure that sequentially estimates both the model parameters and the orientation parameters as follows
\begin{equation} \label{main3d_m}
\bold{m}^+=\argmin_{\bold{m}} \frac{\mu}{2}\|\bold{Gm}-\bold{d}\|_2^2+\frac12\sum_{i=1}^N\| \bold{R}(\bolds{\strike}_i,\bolds{\tilt}_i,\bolds{\dip}_i)(\nabla \bold{m})_i\|_{\bolds{\Sigma}}^2,
\end{equation} 
\begin{argmini}
      {(\bolds{\strike},\bolds{\tilt},\bolds{\dip})\in B}
{\hspace{-0.25cm}\frac12\sum_{i=1}^N\| \bold{R}(\bolds{\strike}_i,\bolds{\tilt}_i,\bolds{\dip}_i)(\tilde{\nabla} \bold{m}^+)_i\|_{\bolds{\Sigma}}^2}
      {\label{main3d_theta}}
      {\begin{pmatrix}
          \bolds{\strike}^+\\
          \bolds{\tilt}^+\\
          \bolds{\dip}^+
      \end{pmatrix}= \hspace{-0.5cm}}
      \breakObjective{+ \frac{\rho_{\strike}}{2}\|\nabla\bolds{\strike}\|_2^2 +\frac{\rho_{\tilt}}{2}\|\nabla\bolds{\tilt}\|_2^2+ \frac{\rho_{\dip}}{2}\|\nabla\bolds{\dip}\|_2^2}.
\end{argmini}
Here $\rho_{\dip},\rho_{\strike},\rho_{\tilt}$ are positive parameters controlling the smoothness of the estimated dip, strike, and tilt fields, and $B=(-\pi/2,\pi/2]^3$. As in the 2D case, $\tilde{\nabla}\bold{m}$ represents a smoothed (regularized) estimate of the model gradient $\nabla \bold{m}$. We achieve this by smoothing the gradient components, $\tilde{\bold{m}}_x=\tilde{\nabla}_x\bold{m}$, $\tilde{\bold{m}}_y=\tilde{\nabla}_y\bold{m}$ and $\tilde{\bold{m}}_z=\tilde{\nabla}_z\bold{m}$, using 3D smooth derivative filters $\bold{h}_x$, $\bold{h}_y$ and $\bold{h}_z$:
\begin{equation} \label{dxz}
\tilde{\bold{m}}_x=\bold{h}_x\star \bold{m},\quad \tilde{\bold{m}}_y=\bold{h}_y \star \bold{m},\quad \tilde{\bold{m}}_z=\bold{h}_z \star \bold{m},
\end{equation} 
with the generalized 3D Hilbert (Riesz) transform filters
\begin{equation} \label{eq1}
\begin{split}
h_x &= -\frac{1}{\pi^2}\frac{x}{(x^2+y^2+z^2)^2},\\
h_y &= -\frac{1}{\pi^2}\frac{y}{(x^2+y^2+z^2)^2},\\
h_z &= -\frac{1}{\pi^2}\frac{z}{(x^2+y^2+z^2)^2}.
\end{split}
\end{equation}
%where $R=\sqrt{x^2+y^2+z^2}$.
\subsection{Solving the Model Subproblem in \eqref{main3d_m}}
Applying the optimality conditions gives
\begin{equation} 
\bold{m}= \left(\mu\bold{G}^T\bold{G}+\bold{D}(\bolds{\strike},\bolds{\tilt},\bolds{\dip})^T\bold{D}(\bolds{\strike},\bolds{\tilt},\bolds{\dip})\right)^{-1}
(\mu\bold{G}^T\bold{d}),
\end{equation}
where $\bold{D}(\bolds{\strike},\bolds{\tilt},\bolds{\dip})$ is the 3D local anisotropic differential operator
{
\begin{equation} \label{D3D}
\bold{D}(\bolds{\strike},\bolds{\dip},\bolds{\tilt}) = \!\!
\begin{pmatrix}
\bold{I}  & \bold{0} & \bold{0}\\
\bold{0} & \sqrt{\delta}\bold{I} & \bold{0}\\
\bold{0} & \bold{0} & \sqrt{\epsilon}\bold{I}
\end{pmatrix}\!\!\!\!
\begin{pmatrix}
\C{\bolds{\tilt}}\C{\bolds{\strike}}  & \C{\bolds{\tilt}}\S{\bolds{\strike}}  & \S{\bolds{\tilt}}\\
%---------------------------------------------------------------------------------------
-\C{\bolds{\dip}} \S{\bolds{\strike}}- \S{\bolds{\dip}}\S{\bolds{\tilt}}\C{\bolds{\strike}} & 
\C{\bolds{\dip}}\C{\bolds{\strike}} -\S{\bolds{\dip}}\S{\bolds{\tilt}}\S{\bolds{\strike}} & \S{\bolds{\dip}}\C{\bolds{\tilt}}\\
%---------------------------------------------------------------------------------------
\S{\bolds{\dip}}\S{\bolds{\strike}} - \C{\bolds{\dip}}\S{\bolds{\tilt}}\C{\bolds{\strike}} &
-\C{\bolds{\dip}}\S{\bolds{\strike}} - \C{\bolds{\dip}}\S{\bolds{\tilt}}\S{\bolds{\strike}}& \C{\bolds{\dip}}\C{\bolds{\tilt}}
\end{pmatrix}\!\!\!\!
\begin{pmatrix}
\nabla_{\!x}\\
\nabla_{\!y}\\
\nabla_{\!z}
\end{pmatrix}.
\end{equation}
}
{Here $C_{\bullet}$ and $S_{\bullet}$ {(with $\bullet$ standing for $\strike,\tilt,\dip$)} are the diagonal matrices with $\cos(\bullet)$ and $\sin(\bullet)$ on their main diagonal.} 
\subsection{Solving the Orientation-Parameters Subproblem in \eqref{main3d_theta}}
Similarly to the 2D case discussed above, after introducing the auxiliary variables $\bold{z}_\bolds{\strike}$, $\bold{z}_\bolds{\tilt}$, $\bold{z}_\bolds{\dip}$ and the constraints $\bold{z}_\bolds{\strike}=\bolds{\strike}$, $\bold{z}_\bolds{\tilt}=\bolds{\tilt}$, $\bold{z}_\bolds{\dip}=\bolds{\dip}$, we employ the ADMM to solve \eqref{main3d_theta} sequentially, with each angle updated independently at each iteration, as follows (see Appendix C):
%{
\begin{subequations}
\begin{align}
\bolds{\strike}^{+} &=\! \bold{H}_{\strike}(\bolds{m}^+,\bolds{\strike},\bolds{\tilt},\bold{\dip})^{-1}\bold{g}_{\strike}(\bolds{m}^+,\bolds{\strike},\bolds{\tilt},\bold{\dip},\bold{z}_{\strike},\bolds{\lambda}_{\strike}), \label{sub_strike3d}\\
\bolds{\tilt}^{+} &=\! \bold{H}_{\tilt}(\bolds{m}^+,\bolds{\strike}^+,\bolds{\tilt},\bold{\dip})^{-1}\bold{g}_{\tilt}(\bolds{m}^+,\bolds{\strike}^+,\bolds{\tilt},\bold{\dip},\bold{z}_{\tilt},\bolds{\lambda}_{\tilt}), \label{sub_tilt3d}\\
\bolds{\dip}^{+} &=\! \bold{H}_{\dip}(\bolds{m}^+\!,\bolds{\strike}^+\!,\bolds{\tilt}^+\!,\bold{\dip})^{-1}\bold{g}_{\dip}(\bolds{m}^+\!,\bolds{\strike}^+\!,\bolds{\tilt}^+\!,\bold{\dip},\bold{z}_{\dip},\bolds{\lambda}_{\dip}), \label{sub_dip3d}\\
\bold{z}_{\strike}^{+} &= \min(\max(\bolds{\strike}^{+}-\bolds{\lambda}_{\strike},-\frac{\pi}{2}),\frac{\pi}{2}),\\
\bold{z}_{\tilt}^{+} &= \min(\max(\bolds{\tilt}^{+}-\bolds{\lambda}_{\tilt},-\frac{\pi}{2}),\frac{\pi}{2}),\\
\bold{z}_{\dip}^{+} &= \min(\max(\bolds{\dip}^{+}-\bolds{\lambda}_{\dip},-\frac{\pi}{2}),\frac{\pi}{2}), \\
\bolds{\lambda}_{\strike}^{+}    &=\bolds{\lambda}_{\strike} - \tau_{\strike} (\bolds{\strike}^{+}-\bold{z}_{\strike}^{+}),\\
\bolds{\lambda}_{\tilt}^{+}    &=\bolds{\lambda}_{\tilt} - \tau_{\tilt} (\bolds{\tilt}^{+}-\bold{z}_{\tilt}^{+}),\\
\bolds{\lambda}_{\dip}^{+}  &=\bolds{\lambda}_{\dip} - \tau_{\dip} (\bolds{\dip}^{+}-\bold{z}_{\dip}^{+}),
\end{align}
\end{subequations}
%}
where $\bullet$ {(which stands for $\strike,\tilt,\dip, z_{\strike},z_{\tilt},z_{\dip},\lambda_{\strike},\lambda_{\tilt},\lambda_{\dip}$)} are the values at the current iteration, $\bullet^{+}$ are the updated values, and $\tau_{\strike}$, $\tau_{\tilt}$, $\tau_{\dip}$ are penalty parameters. The Hessian matrices $\bold{H}_{\strike},\bold{H}_{\tilt},\bold{H}_{\dip}$ and right-hand-side vectors $\bold{g}_{\strike},\bold{g}_{\tilt},\bold{g}_{\dip}$ are defined in equations \ref{Hess3d} and \ref{g_strike}, respectively. 

\subsection{Selection of the regularization parameters}
The procedure outlined for selecting regularization parameters in the 2D case can be extended to the 3D algorithm. The parameter $\mu$ in \eref{main3d_m} can be dynamically adjusted using \eref{alpha_update}. In each iteration, the regularization parameters $\rho_{\bullet}$ and $\tau_{\bullet}$, where $\bullet$ represents $\strike$, $\tilt$, and $\dip$, are set to be proportional to the maximum value of the diagonal of the associated Gauss-Newton Hessian $\bold{J}_{\bullet}^T\bold{J}_{\bullet}$, with the scale parameters $\epsilon_1$ and $\epsilon_2$ (see equations \ref{rho} and \ref{tau}). For the numerical examples in this paper, we use $\epsilon_1=10$ and $\epsilon_2=0.1$.

The proposed algorithm for 3D problems is summarized in Algorithm \ref{Algorithm3D}. %Note that, to follow the procedure just described, one should take $K=1$. 

\begin{algorithm}[!h]
{%\footnotesize
\vspace{.2cm}
 \caption{Structural-orientation parameter estimation and local anisotropic regularization in 3D space.}  \label{Algorithm3D}
 Inputs:  data $\bold{d}$ and forward operator $\bold{G}$, and initial penalty parameter $\mu$, $\epsilon_1$ and $\epsilon_2$.\\ 
 Set  $\bolds{\strike}=\bolds{\tilt}=\bolds{\dip}=\bold{z}_{\strike}=\bold{z}_{\tilt}=\bold{z}_{\dip}=\bolds{\lambda}_{\strike}=\bolds{\lambda}_{\tilt}=\bolds{\lambda}_{\dip}=\bold{0}$.   \\ \vspace{2mm} 
 \While{\text{a stopping criterion is {not} satisfied} }
 {
%--------------------------------------------------------------------------------------
%\%            update the model \\
\vspace{2mm}
$\bold{m} \leftarrow (\mu\bold{G}^T\bold{G}+\bold{D}(\bolds{\strike},\bolds{\tilt},\bolds{\dip})^T\bold{D}(\bolds{\strike},\bolds{\tilt},\bolds{\dip}))^{-1}(\mu\bold{G}^T\bold{d})$ \\  \vspace{2mm} 
%--------------------------------------------------------------------------------------
\For{\text{$k=1,...,K$}}{
%--------------------------------------------------------------------------------------
%\% update the strike angle\\
$\bolds{\strike} \leftarrow \bold{H}_{\strike}(\bolds{m},\bolds{\strike},\bolds{\tilt},\bolds{\dip})^{-1}\bold{g}_{\strike}(\bolds{m},\bolds{\strike},\bolds{\tilt},\bolds{\dip},\bold{z}_{\strike},\bolds{\lambda}_{\strike})$\\ \vspace{2mm} 
%--------------------------------------------------------------------------------------
%\% update the dip angle\\
$\bolds{\tilt} \leftarrow \bold{H}_{\tilt} (\bolds{m},\bolds{\strike},\bolds{\tilt},\bolds{\dip})^{-1}\bold{g}_{\tilt}(\bolds{m},\bolds{\strike},\bolds{\tilt},\bold{z}_{\tilt},\bolds{\lambda}_{\tilt})$\\ \vspace{2mm} 
%--------------------------------------------------------------------------------------
%\% update the tilt angle\\
$\bolds{\dip} \leftarrow \bold{H}_{\dip} (\bolds{m},\bolds{\strike},\bolds{\tilt},\bolds{\dip})^{-1}\bold{g}_{\dip}(\bolds{m},\bolds{\strike},\bolds{\tilt},\bolds{\dip},\bold{z}_{\dip},\bolds{\lambda}_{\dip})$\\ \vspace{2mm} 
%--------------------------------------------------------------------------------------
%\% update the auxiliary variable\\
 $\bold{z}_{\strike}   \leftarrow\min(\max(\bolds{\strike}-\bolds{\lambda}_{\strike},-\frac{\pi}{2}),\frac{\pi}{2})$ \\ \vspace{2mm} 
$\bold{z}_{\tilt}  \leftarrow \min(\max(\bolds{\tilt}-\bolds{\lambda}_{\tilt},-\frac{\pi}{2}),\frac{\pi}{2})$ \\ \vspace{2mm} 
$\bold{z}_{\dip}   \leftarrow \min(\max(\bolds{\dip}-\bolds{\lambda}_{\dip},-\frac{\pi}{2}),\frac{\pi}{2})$ \\ \vspace{2mm} 
%--------------------------------------------------------------------------------------
%\% update the Lagrange multipliers\\
$\bolds{\lambda}_{\strike}  \leftarrow\bolds{\lambda}_{\strike} - \tau_{\strike} (\bolds{\strike}-\bold{z}_{\strike})$\\ \vspace{2mm} 
$\bolds{\lambda}_{\tilt} \leftarrow\bolds{\lambda}_{\tilt} - \tau_{\tilt} (\bolds{\tilt}-\bold{z}_{\tilt})$\\ \vspace{2mm} 
 $\bolds{\lambda}_{\dip} \leftarrow \bolds{\lambda}_{\dip} - \tau_{\dip} (\bolds{\dip}-\bold{z}_{\dip})$\\ \vspace{2mm} 
 }
%--------------------------------------------------------------------------------------
%\% update the data fit parameter\\
$\mu\leftarrow\dfrac{2\|\bold{Gm}-\bold{d}\|_2^2}{\|\bold{Gm}-\bold{d}\|_2^2+\varepsilon}\mu$
}
}
Outputs: model $\bold{m}$, tilt angles $\theta$, strike angles $\phi$, dip angles $\psi$
\end{algorithm}

%-----------------------------------------------------------------------------
%-----------------------------------------------------------------------------------
%--------------------------------------------- 3D algorithm
%-----------------------------------------------------------------------------------

%%%%%%%%%%%%%%%%%%%%%%%%%%%%%%%%%%%%%%%%%%%%%%%%%%%%%%%%%%%%%%%%%%%%%%%%%%%%%%%%%%%%%%%%%%%%%%%%
\begin{figure}
\center
\includegraphics[width=0.7\columnwidth]{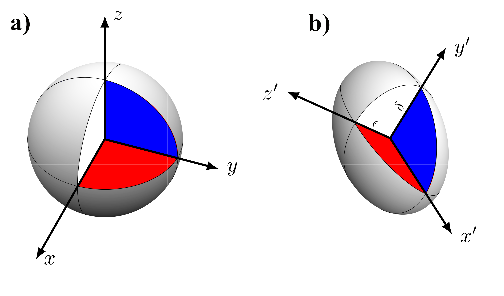}
\caption{Ellipsoid representation of (a) isotropic and (b) anisotropic regularization in a 3D space.  }
\label{ellipsoinds}
\end{figure}

\section{Numerical examples} \label{NumEx}
In this section, we conduct a comprehensive experimental evaluation of the proposed local anisotropic regularization method coupled with structural-orientation parameter estimation, employing a variety of detailed 2D and 3D examples. 
%To uphold consistency throughout our numerical experiments, we have thoughtfully chosen specific parameter values.
%For our 2D examples, we have set $\epsilon_1=1$ and $\epsilon_2=0.1$ in equations \ref{rho} and \ref{tau}, along with $\epsilon=0.01$ for the anisotropy parameter. %On the other hand, 
%For our 3D examples, the values for $\rho$ and $\tau$ are determined as per equations \ref{rho} and \ref{tau}, featuring $\epsilon_1=10$ and $\epsilon_2=0.1$. Additionally, the anisotropy parameters are chosen as $\delta=0.1$ and $\epsilon=0$. 

{
In our algorithms, there are two main iterations: the outer loop, which alternates between estimating the model and orientation parameter(s), and the inner loop, which iterates over the orientation parameter(s) for a given model. The outer loop continues until the change in successive model estimates and orientation parameter(s) estimate falls below a predefined threshold. The inner loop is characterized by the index $k$ in Algorithms \ref{Algorithm2D} and \ref{Algorithm3D}, and the maximum number of these iterations is denoted by $K$. For all the examples presented in this paper, we set $K=1$.
In each iteration of the algorithms, a set of linear systems associated with the model and with the orientation parameter(s) need to be solved. These systems are solved either by direct solvers for small to medium-sized problems or by an iterative solver like conjugate-gradient (CG) for larger problems. Our algorithms are not highly sensitive on the number of CG iterations, but these should nonetheless be appropriately set to ensure accurate solutions for the associated linear systems. In the 3D examples presented in this section, we stop the CG iteration either after reaching a maximum of 200 iterations or when the relative residual error falls below $10^{-4}$.
}

\subsection{2D examples: Simultaneous denoising and tilt estimation}
\subsubsection{Synthetic data}
We start our investigation by addressing simultaneous denoising (with different signal-to-noise ratios) and tilt estimation, 
representing a crucial test for evaluating the performance of the proposed algorithm. Our choice of seismic data, illustrated in Figure \ref{signals_slope}a, comprises a synthetic model representing sedimentary layers featuring a plane unconformity and a curved fault, following the design by \citet{Fomel_2002_AOP}. This model consists of 512 traces, each with 512 samples. To process this signal, we employ Algorithm \ref{Algorithm2D} with $\bold{G}=\bold{I}$ (the identity matrix).

In the initial phase, we introduce no noise to the model, thereby setting a very low noise level $\varepsilon$ to determine the regularization parameter for model reconstruction, as displayed in Figure \ref{alpha_update}. Figure \ref{signals_slope}b shows the estimated tilt field. {The tilt angle is accurately determined at most points in the signal, except at positions corresponding to unconformities and faults, where the gradient is undefined. The value of the tilt angle at these positions is estimated based on a combination of neighboring points, the initial guess of the tilt, and the regularization applied during the estimation process. Figure \ref{signals_slope}c shows the corresponding convergence curve.}
These results demonstrate the algorithm's remarkable speed and stability during convergence.

To assess the algorithm's robustness, we introduce noise to the model {shown} in Figure \ref{signals_slope}a, targeting signal-to-noise ratios (S/N) of 20 dB, 10 dB, and 1 dB. After generating noisy data (Figure \ref{noise_level}a-c), employing Algorithm \ref{Algorithm2D} with a given noise level $\varepsilon$, we compute the denoised data (Figure \ref{noise_level}d-f), and the estimated tilt fields (Figure \ref{noise_level}g-i). For further insight, Figure \ref{noise_level_params}a-c show a direct comparison of clean, noisy, and estimated signals for a specific trace segment in the middle of the section. {We observe that, as the noise level increases, both the denoised signal and the estimated tilt become smoother; however, the continuous parts of the signal are less affected, thanks to the structure-oriented smoothing. This smoothing effect is more pronounced at unconformities and faults.}
Figures \ref{noise_level_params}d-f show the corresponding convergence curves for the misfit function, penalty parameter $\mu$, and regularization functional versus iteration, respectively. %These results highlight the algorithm's robustness and stability, even under challenging noise conditions.
Although, in Figure \ref{noise_level_params}d the misfit value (defined as $1/2\|\bold{d}-\bold{m}\|_2^2$) is reported for coherence with the metrics used  later in this section to evaluate the performance of our method on other test problems, its final value in terms of S/N is 27 dB (for input S/N 20 dB), 22 dB (for input S/N 10 dB), 18 dB (for input S/N 1 dB).
In Figure \ref{MSE_vs_SNR}, the mean-squared error (MSE) is plotted for both the denoised signal and the estimated tilt field at different noise levels, utilizing the clean signal and its associated tilt field as reference benchmarks. In general, looking at Figures \ref{noise_level}-\ref{MSE_vs_SNR}, we can conclude that our new algorithm exhibits high stability against varying noise levels.

As stated at the beginning of this section, the results presented so far are obtained using the parameters $\epsilon_1=1$ and $\epsilon_2=0.1$ when determining the regularization parameters $\rho$ and $\tau$ through equations \ref{rho} and \ref{tau}. To assess the algorithm's sensitivity to these parameters, we conducted a test using a noisy signal with an S/N of 10 dB, as illustrated in the middle column of Figure \ref{noise_level}. Algorithm \ref{Algorithm2D} was run for simultaneous denoising and tilt estimation, employing different values for $\epsilon_1$ and $\epsilon_2$. Figure \ref{Figure09} shows the MSE of the denoised signals. We observe that the algorithm is stable for a wide range of these parameters with greater sensitivity to the choice of $\epsilon_1$ than $\epsilon_2$. This discrepancy arises from the fact that $\epsilon_2$ governs the value of $\tau$, imposing constraints on the tilt angles within the range of $-\pi/2$ to $\pi/2$, but $\epsilon_1$ enforces the degree of smoothness in the tilt field through the regularization parameters $\rho$. Although the corresponding reconstructions are not reported here, we observe that enforcing varying degrees of smoothness of the tilt field primarily affects the quality of estimate around the discontinuities.

\subsubsection{Real data}
Expanding our investigation to real seismic post-stack data, {shown} in Figure \ref{real_data_stack}a, we consider complex structures associated with discontinues features.
The data are from the North Viking Graben area of the North Sea and were published by ExxonMobil \citep{Keys_1998_CSI}. An
interpretation of this data can be found in \citet{Madiba_2003_PII}.
%{The data corresponds to the North Viking Graben area of the North Sea and were published by ExxonMobil \citep{Keys_1998_CSI}; interpretation of the data can be found in \citep{Madiba_2003_PII}.
The selected data region includes a complex toplap reflection package, including steeply dipping strata terminating against an overlying discontinuity surface, alongside nearly horizontal layers above. Such complex geological structures pose significant challenges for conventional isotropic denoising algorithms, which are typically equally smooth along every direction.
We apply Algorithm \ref{Algorithm2D} to denoise the seismic section, determining the noise level $\varepsilon$ through trial and error-verified by observing the estimated noise becoming free of coherent energy. Figures \ref{real_data_stack}b-d show the denoised section, the estimated noise, and the estimated tilt angle, respectively. 
The results show that, by employing anisotropic regularizers that act along the progressively estimated tilt field, the denoising is effective, preserving reflections while removing a significant portion of incoherent noise, as evident in Figure \ref{real_data_stack}b. To provide further insight, Figure \ref{real_data_params} tracks the evolution of the misfit function, penalty parameter $\mu$, and regularization functional during iterations. 
These curves show the stability of the denoising algorithm in handling real seismic data containing complex structures such as discontinuities and steeply dipping layers. By effectively preserving discontinuities and reflections while suppressing noise, the algorithm enhances the interpretability of the data.
This comprehensive analysis validates the ability of the algorithm to produce high-quality denoised sections, even when dealing with real seismic data featuring complex structures.
}
%%%%%%%%%%%%%%%%%%%%%%%%%%%%%%%%%%%%%%%%%%%%%%%%%%%%%%%%%%%%%%%%%%%%%%%%%%%%%%%%%%%%%%%%%%%%%%%%
\begin{figure}
\center
\includegraphics[width=1\columnwidth]{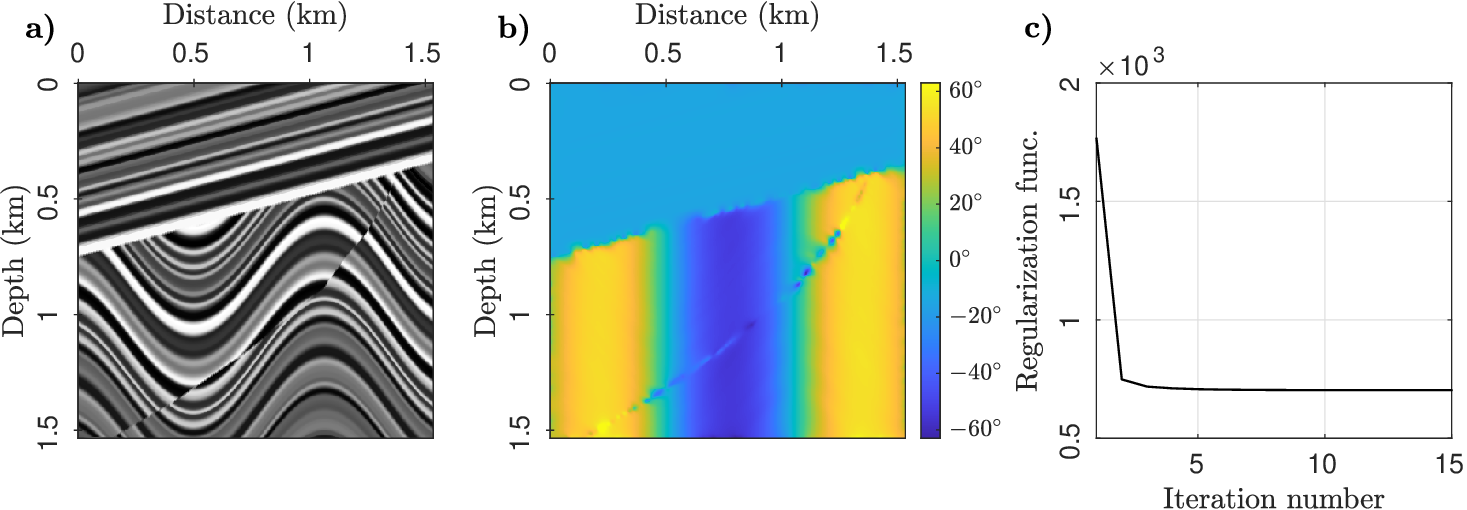}
\caption{(a) Synthetic sedimentary model. (b) Tilt field estimation achieved through the proposed method. (c) Corresponding convergence curve depicting the algorithm's stability and convergence characteristics. }
\label{signals_slope}
\end{figure}
%%%%%%%%%%%%%%%%%%%%%%%%%%%%%%%%%%%%%%%%%%%%%%%%%%%%%%%%%%%%%%%%%%%%%%%%%%%%%%%%%%%%%%%%%%%%%%%%
\begin{figure}
\center
\includegraphics[width=1\columnwidth]{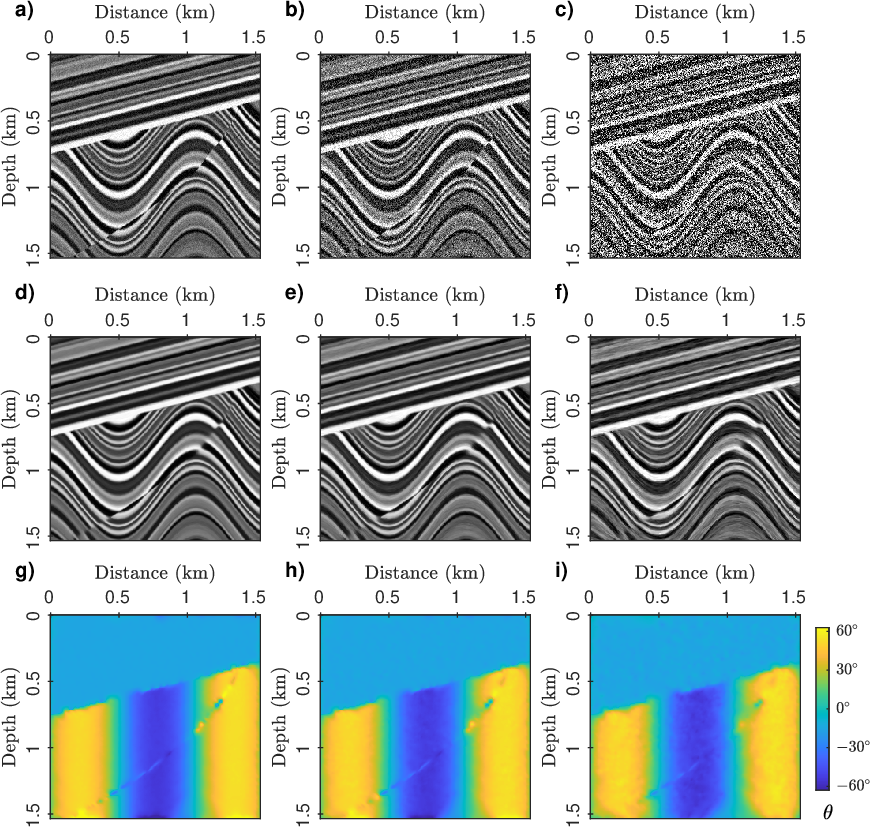}
\caption{(a-c) Noisy signals with different signal-to-noise ratios: (a) S/N = 20 dB, (b) S/N = 10 dB, and (c) S/N = 1 dB. (d-f) Corresponding signals estimated using the anisotropic Tikhonov filter. (g-i) Corresponding estimated tilt fields. 
}
\label{noise_level}
\end{figure}

%%%%%%%%%%%%%%%%%%%%%%%%%%%%%%%%%%%%%%%%%%%%%%%%%%%%%%%%%%%%%%%%%%%%%%%%%%%%%%%%%%%%%%%%%%%%%%%%
\begin{figure}
\center
\includegraphics[width=1\columnwidth]{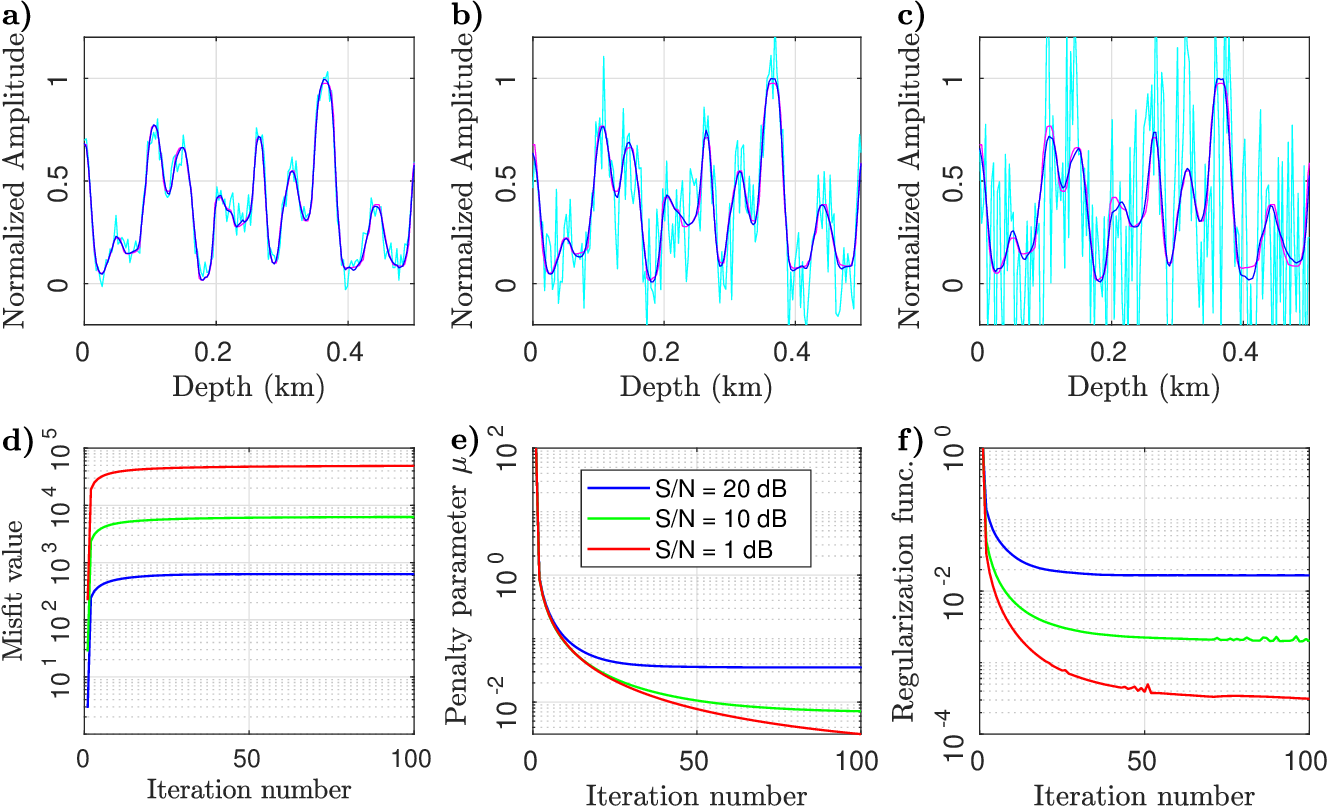}
\caption{ (a-c) A section (0 to 0.5 km) of the trace at a distance of 0.75 km for the denoising examples presented in Figure \ref{noise_level}. Shown are (a) S/N = 20 dB, (b) S/N = 10 dB, and (c) S/N = 1 dB scenarios. In these panels, the true signal, noisy signal, and denoised signal are represented by magenta, cyan, and blue lines, respectively. (d-f) The evolution of key metrics: (d) Misfit function, (e) Penalty parameter, and (f) Regularization functional as a function of iteration.}
\label{noise_level_params}
\end{figure}
%%%%%%%%%%%%%%%%%%%%%%%%%%%%%%%%%%%%%%%%%%%%%%%%%%%%%%%%%%%%%%%%%%%%%%%%%%%%%%%%%%%%%%%%%%%%%%%%
\begin{figure}
\center
\includegraphics[width=0.5\columnwidth]{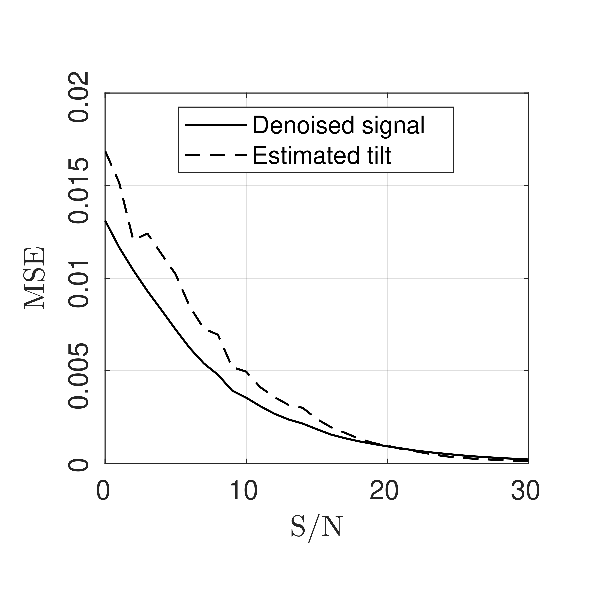}
\caption{Performance evaluation of Algorithm \ref{Algorithm2D} for simultaneous denoising and tilt estimation for different noise levels. The MSE is calculated for both the denoised signal and the estimated tilt field, using the clean signal and its associated tilt field as references for comparison. }
\label{MSE_vs_SNR}
\end{figure}

%%%%%%%%%%%%%%%%%%%%%%%%%%%%%%%%%%%%%%%%%%%%%%%%%%%%%%%%%%%%%%%%%%%%%%%%%%%%%%%%%%%%%%%%%%%%%%%%
\begin{figure}
\center
\includegraphics[width=1\columnwidth]{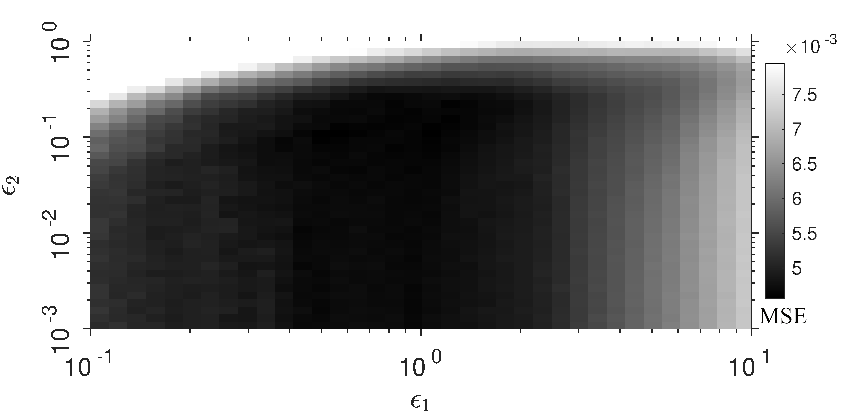}
\caption{The MSE of Algorithm \ref{Algorithm2D} for denoising a noisy signal with a S/N of 10 dB (shown in the middle column of Figure \ref{noise_level}), with different values of the parameters $\epsilon_1$ and $\epsilon_2$ appearing in formulas \ref{rho} and \ref{tau} for the determination of the regularization parameter for tilt field estimation, and for the penalty parameters in ADMM.}
\label{Figure09}
\end{figure}

%%%%%%%%%%%%%%%%%%%%%%%%%%%%%%%%%%%%%%%%%%%%%%%%%%%%%%%%%%%%%%%%%%%%%%%%%%%%%%%%%%%%%%%%%%%%%%%%
\begin{figure}
\center
\includegraphics[width=1\columnwidth]{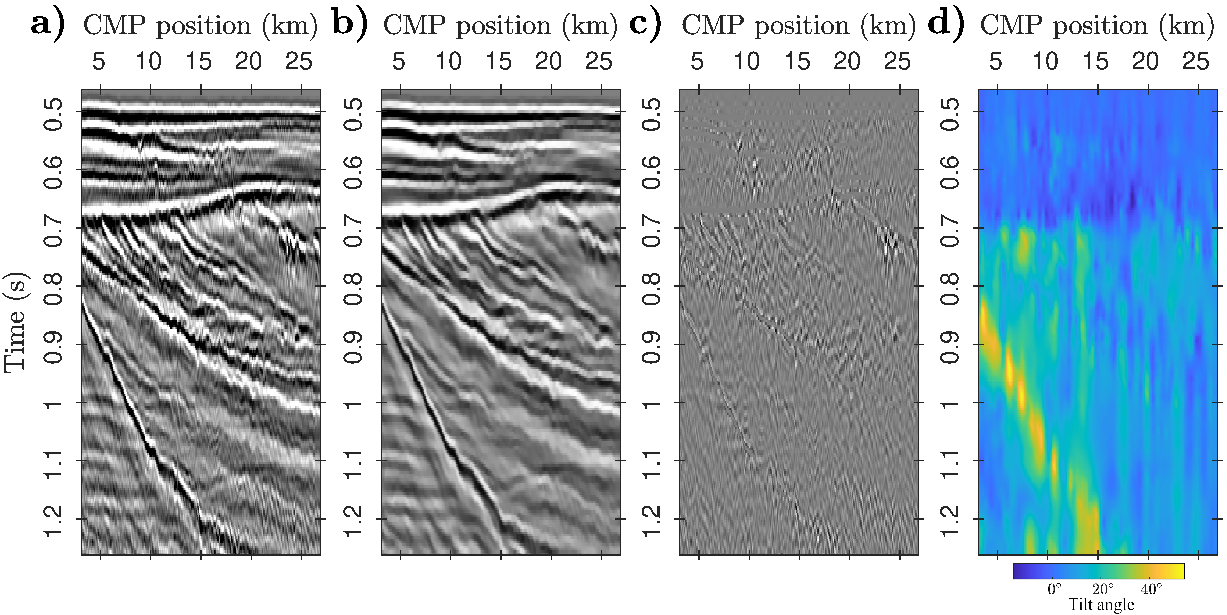}
\caption{(a) Stack section of seismic field data. (b) Denoised section resulting from the application of Algorithm \ref{Algorithm2D}. (c) Section depicting the estimated noise. (d) Tilt field estimation. The progression of relevant optimization parameters is presented in Figure \ref{real_data_params}.}
\label{real_data_stack}
\end{figure}

%%%%%%%%%%%%%%%%%%%%%%%%%%%%%%%%%%%%%%%%%%%%%%%%%%%%%%%%%%%%%%%%%%%%%%%%%%%%%%%%%%%%%%%%%%%%%%%%
\begin{figure}
\center
\includegraphics[width=1\columnwidth]{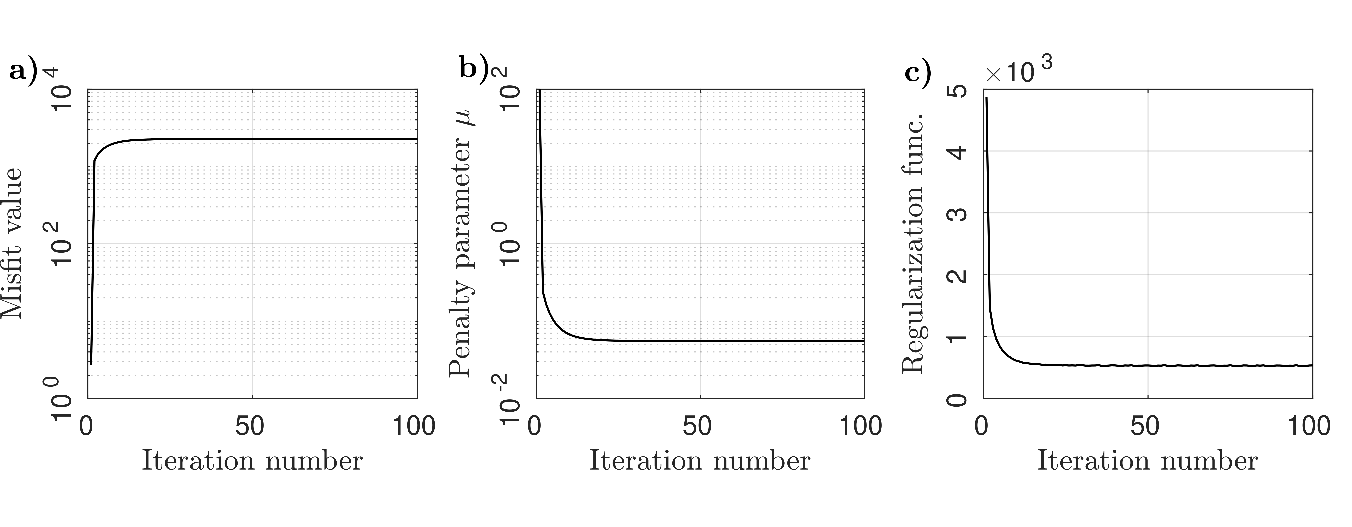}
\caption{The evolution of crucial optimization parameters during the denoising process of the real seismic data presented in Figure \ref{real_data_stack}. (a) shows the misfit function, (b) shows the penalty parameter, and (c) shows the regularization functional as a function of iteration.}
\label{real_data_params}
\end{figure}

%%%%%%%%%%%%%%%%%%%%%%%%%%%%%%%%%%%%%%%%%%%%%%%%%%%%%%%%%%%%%%%%%%%%%%%%%%%%%%%%%%%%%%%%%%%%%%%%

\subsection{3D examples}
\subsubsection{3D synthetic Qdome model}
\begin{figure}[htb!] 
\begin{center}
\begin{tabular}{cccc}
\hspace{-0cm}\small{\textbf{a)}} & \hspace{-0cm}\small{\textbf{b)}}  \\
\hspace{-0cm}\includegraphics[scale=0.4]{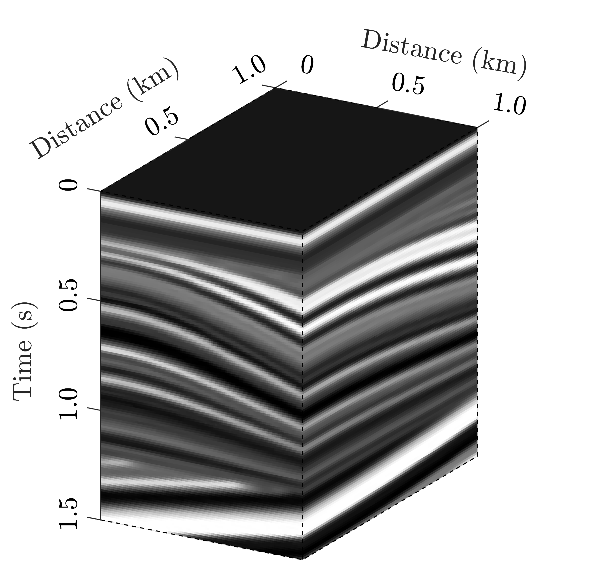} &
\hspace{-0cm}\includegraphics[scale=0.4]{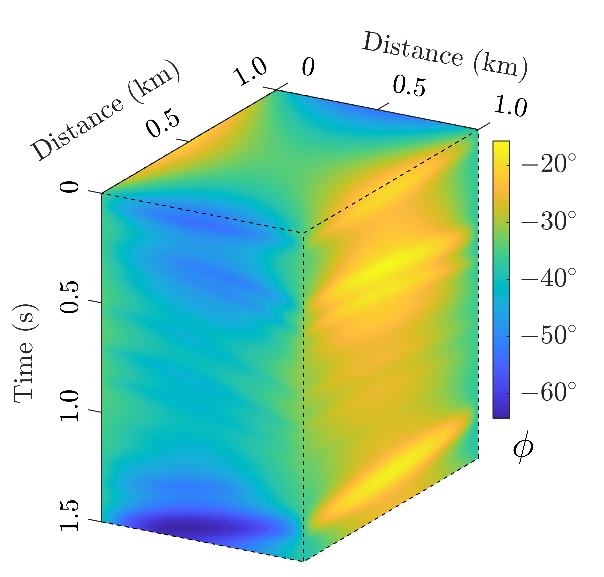} \\
\hspace{-0cm}\small{\textbf{c)}} & \hspace{-0cm}\small{\textbf{d)}}  \\
\hspace{-0cm}\includegraphics[scale=0.4]{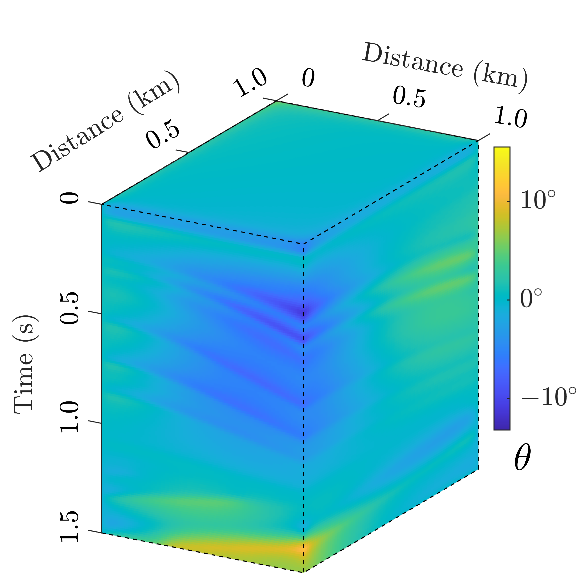}&
\hspace{-0cm}\includegraphics[scale=0.4]{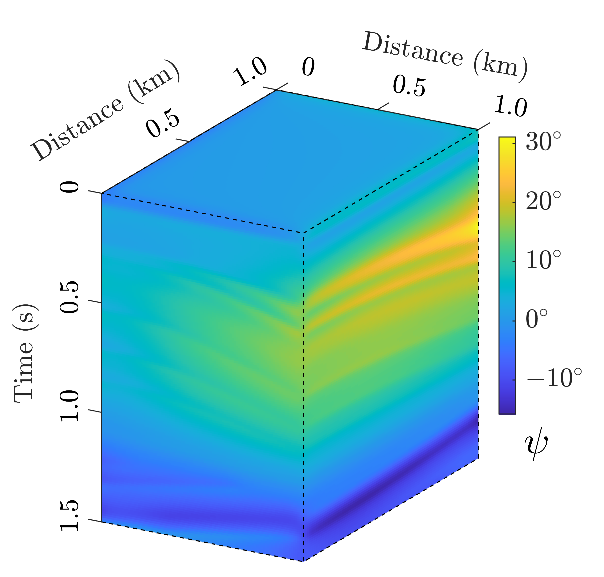} 
\end{tabular} 
\caption{ (a) 3D synthetic Qdome model. (b-d) Estimated orientation parameters, including (b) strike angle $\bolds{\strike}$, (c) tilt angle $\bolds{\theta}$, and (d) dip angle $\bolds{\dip}$.}
\label{qdome_angles}
\end{center}
\end{figure}

In our examination of a 3D synthetic Qdome model, following \citet{Claerbout_19933_LMA} and featuring complex structural elements such as curved events and faults (Figure \ref{qdome_angles}a), we aim to evaluate the effectiveness of the proposed 3D algorithm in simultaneously denoising the data while estimating the stuctural orientation parameters, including dip, strike, and tilt. The model dimensions are 100 (inline traces) $\times$ 100 (xline traces) $\times$ 150 (time samples). For estimating the orientation parameters, we apply Algorithm \ref{Algorithm3D} {with the anisotropy parameters $\delta=0.1$ and $\epsilon=0$.}

\begin{figure}
\center
\includegraphics[width=0.5\columnwidth]{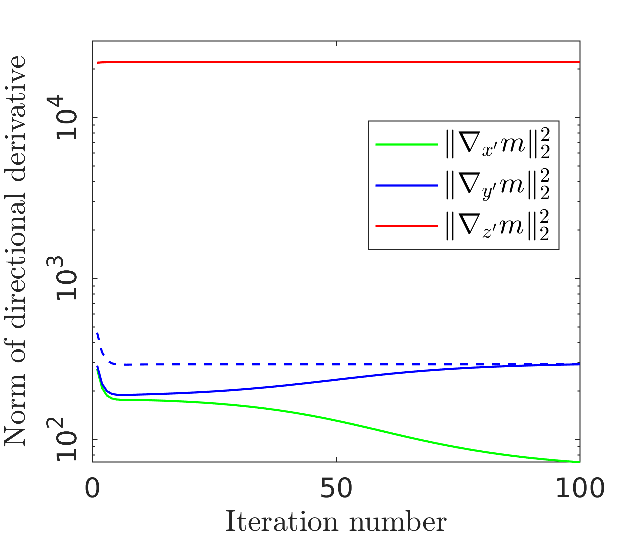}
\caption{3D synthetic Qdome model. Solid lines depict the progression of the norm of directional derivatives during the generation of Figure \ref{qdome_angles} using Algorithm \ref{Algorithm3D}. Dashed lines represent the norm of derivatives when fully minimizing $\|\nabla_{\!\bold{x}'(\strike,\tilt)}\bold{m}\|_2^2$ at each iteration prior to updating the dip angle.}
\label{qdome_angles_ndrives}
\end{figure}

In a first instance, we do not add any noise to the model so that, given the clean nature of the model, we skip the model estimation subproblem (line 4) within Algorithm \ref{Algorithm3D}. This choice enabled us to evaluate the algorithm's efficacy in denoising scenarios where the clean signal is not available, and orientation parameters need to be estimated simultaneously during the denoising process.
To solve the linear systems with coefficient matrices $H_{\bullet}$ appearing in equations \ref{sub_strike3d}-\ref{sub_dip3d}, with $\bullet=\strike,\tilt,\dip$, we employ the preconditioned CG iteration with diagonal preconditioner. 
Figures \ref{qdome_angles}b-\ref{qdome_angles}d illustrate the obtained strike, tilt, and dip angles after 100 iterations of Algorithm \ref{Algorithm3D}. 
Examining the seismic signal in Figure \ref{qdome_angles}a, we observe a downward bending of layers around 0.5 s along the positive spatial axes. Subsequently, the layers flatten around 1.1 s, followed by a reversal in bending direction down to 1.5 s. This characteristic behavior aligns well with the estimated tilt, as shown in Figure \ref{qdome_angles}c, and dip angles in Figure \ref{qdome_angles}d, with the dip angles slightly higher than the tilt. Additionally, it is noteworthy that, in this model, the layers' strike, as depicted in Figure \ref{qdome_angles}b, is quite high, ranging approximately between $-60^\circ$ and $-20^\circ$. 
Figure \ref{qdome_angles_ndrives} provides a graphical representation of the norm of directional derivatives. To emphasize the performance of the alternating direction minimization, we readdress the problem, focusing solely on the dip angle, assuming that the strike and tilt angles are known in advance (see Figures \ref{qdome_angles}b-\ref{qdome_angles}c). The norm of derivatives in directions $y'$ and $z'$, represented by dashed lines in Figure \ref{qdome_angles_ndrives}, highlights the convergence to the same solution, albeit with a difference in the convergence rate.

\begin{figure}[htb!] 
\begin{center}
\begin{tabular}{ccc}
\hspace{-0cm}\small{\textbf{a)}} & \hspace{-0cm}\small{\textbf{b)}} \\
\hspace{-0cm}\includegraphics[scale=0.4]{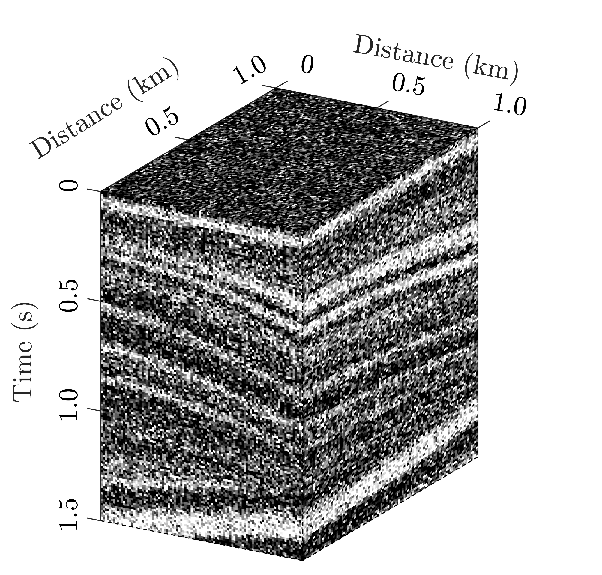} &
\hspace{-0cm}\includegraphics[scale=0.4]{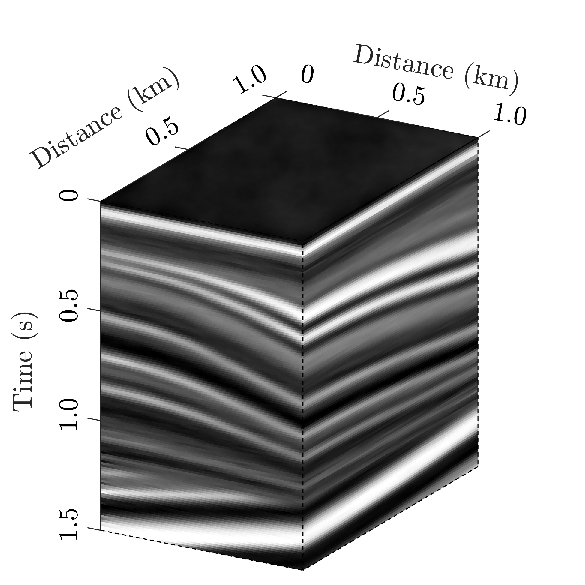} \\ 
\hspace{-0cm}\small{\textbf{c)}} & \hspace{-0cm}\small{\textbf{d)}}\\
\hspace{-0cm}\includegraphics[scale=0.4]{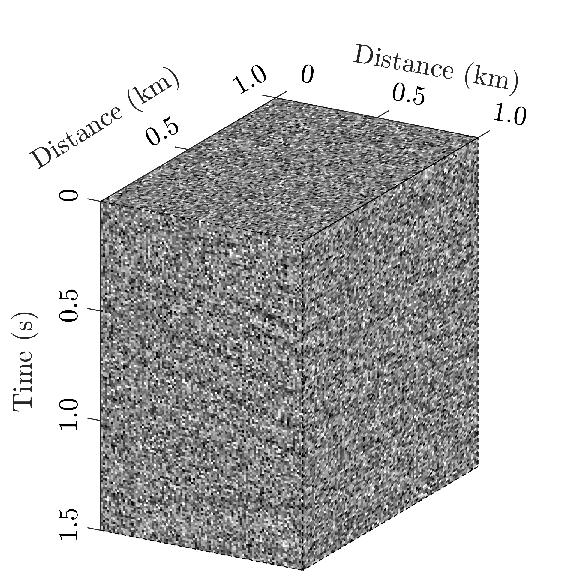} &
\hspace{-0cm}\includegraphics[scale=0.4]{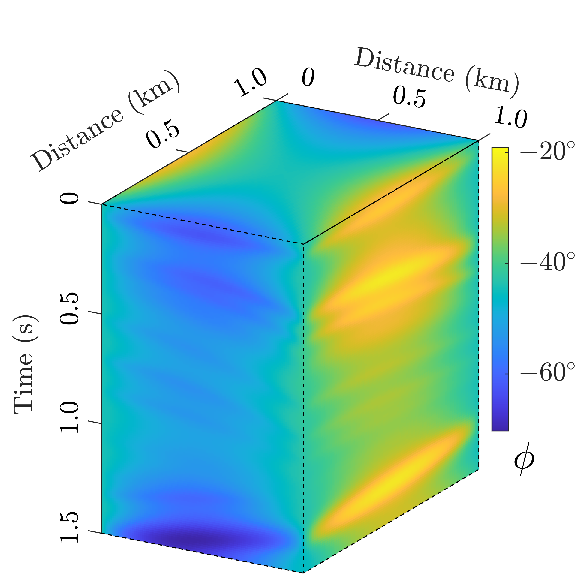} \\
 \hspace{-0cm}\small{\textbf{e)}} & \hspace{-0cm}\small{\textbf{f)}}  \\
\hspace{-0cm}\includegraphics[scale=0.4]{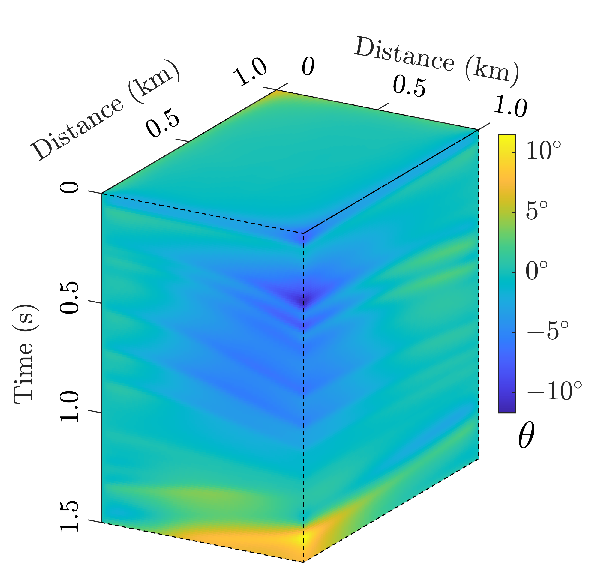} &
\hspace{-0cm}\includegraphics[scale=0.4]{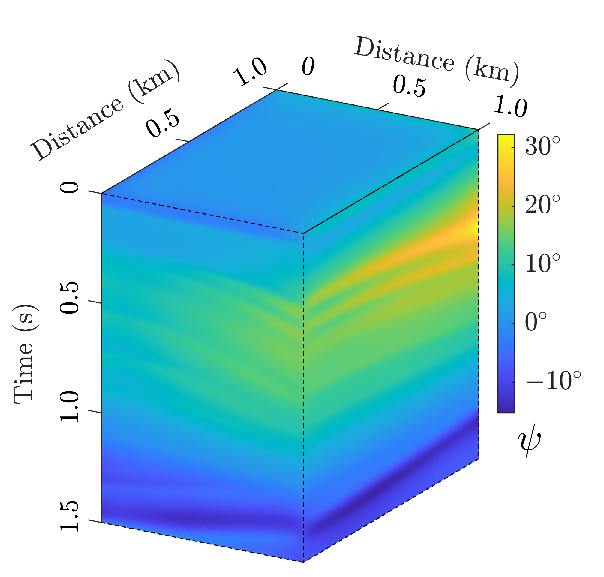}
\end{tabular} 
\caption{ (a) Noisy 3D Qdome model. (b) Denoised model obtained through anisotropic Tikhonov filtering. (c) Estimated noise in the model. (d-f) Estimated orientation parameters, including (d) strike angle $\bolds{\strike}$, (e) tilt angle $\bolds{\theta}$, and (f) dip angle $\bolds{\dip}$. The evolution of relevant optimization parameters is presented in Figure \ref{qdome_denoise_parames}.}
\label{qdome_denoise}
\end{center}
\end{figure}

\begin{figure}
\center
\includegraphics[width=0.75\columnwidth]{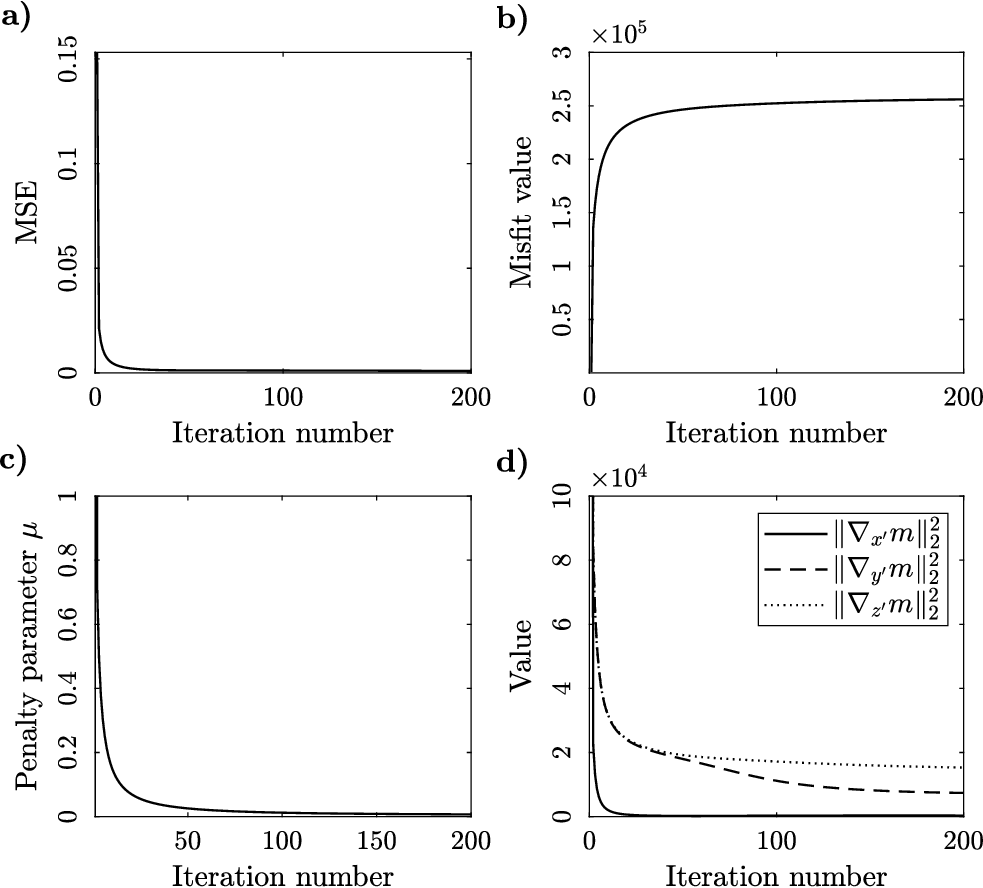}
\caption{The evolution of key optimization parameters during the denoising process for the model {shown} in Figure \ref{qdome_denoise}b. Illustrated are (a) MSE, (b) the misfit function, (c) the penalty parameter $\mu$, and (d) the norm of directional derivatives in the denoised model.}
\label{qdome_denoise_parames}
\end{figure}

{In a second instance, }to make this problem more challenging, we add random noise with S/N = 1 dB to the model, as seen in Figure \ref{qdome_denoise}a. Employing Algorithm \ref{Algorithm3D} and assuming knowledge of the noise level $\varepsilon$, we denoise the data effectively. Figures \ref{qdome_denoise}b and \ref{qdome_denoise}c present the denoised data and the noise that was successfully removed by the anisotropic filter after 200 iterations. Remarkably, the algorithm accurately attenuates the noise while preserving the intricate structural details within the data.
{The estimated strike, tilt, and dip angles are shown in Figures \ref{qdome_denoise}d and \ref{qdome_denoise}f. A direct comparison with the corresponding angles derived from clean data (Figures \ref{qdome_angles}b-\ref{qdome_angles}d) shows the algorithm's stability in the presence of noise in the input data. To provide insights into the algorithm's convergence during the 3D denoising process illustrated in Figure \ref{qdome_denoise}, 
Figure \ref{qdome_denoise_parames} presents the evolution of various parameters across iterations.}

%%%%%%%%%%%%%%%%%%%%%%%%%%%%%%%%%%%%%%%%%%%%%%%%%%%%%%%%%%%%%%%%%%%%%%%%%%%%%%%%%%%%%%%%%%%%%%%%
%%%%%%%%%%%%%%%%%%%%%%%%%%%%%%%%%%%%%%%%%%%%%%%%%%%%%%%%%%%%%%%%%%%%%%%%%%%%%%%%%%%%%%%%%%%%%%%%

%%%%%%%%%%%%%%%%%%%%%%%%%%%%%%%%%%%%%%%%%%%%%%%%%%%%%%%%%%%%%%%%%%%%%%%%%%%%%%%%%%%%%%%%%%%%%%%%
%%%%%%%%%%%%%%%%%%%%%%%%%%%%%%%%%%%%%%%%%%%%%%%%%%%%%%%%%%%%%%%%%%%%%%%%%%%%%%%%%%%%%%%%%%%%%%%%

\subsubsection{3D field Parihaka data}
\begin{figure}[htb!] 
\begin{center}
\begin{tabular}{cccc}
\hspace{-0cm}\small{\textbf{a)}} & \hspace{-0cm}\small{\textbf{b)}}  \\
\hspace{-0cm}\includegraphics[scale=0.4]{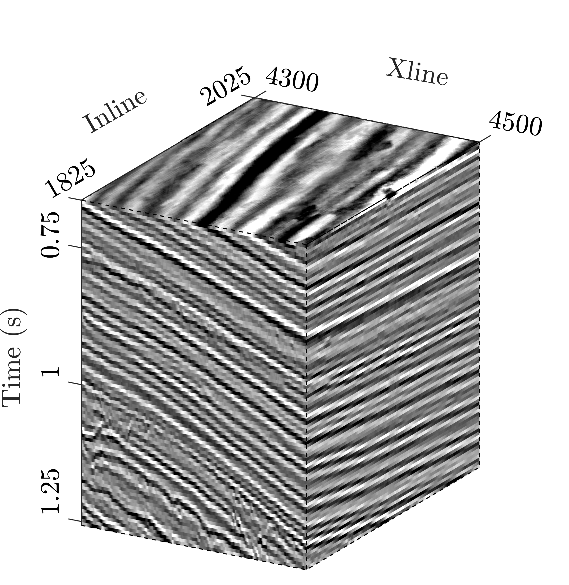} &
\hspace{-0cm}\includegraphics[scale=0.4]{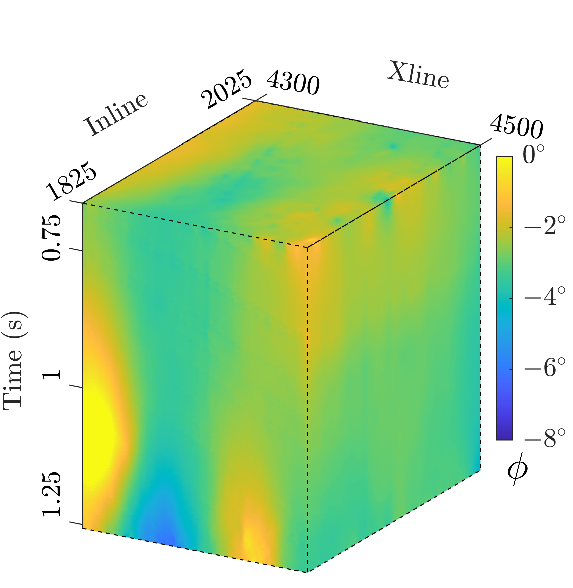} \\
\hspace{-0cm}\small{\textbf{c)}} & \hspace{-0cm}\small{\textbf{d)}}  \\
\hspace{-0cm}\includegraphics[scale=0.4]{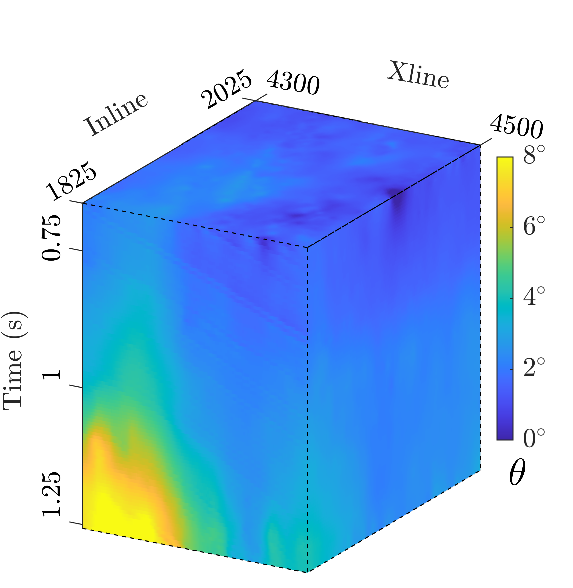} &
\hspace{-0cm}\includegraphics[scale=0.4]{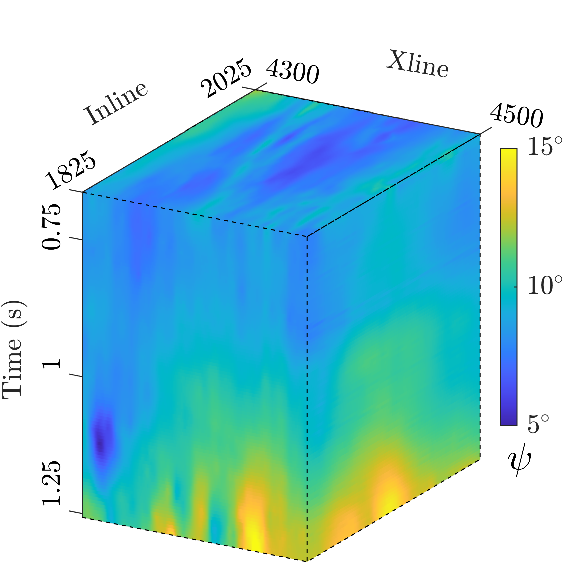}
\end{tabular} 
\caption{(a) 3D Parihaka field data. (b-d) estimated orientation parameters (b) strike angle $\bolds{\strike}$, {(c)} tilt angle $\bolds{\tilt}$, and (d) dip angle $\bolds{\dip}$.}
\label{Parihaka_angles}
\end{center}
\end{figure}

\begin{figure}
\center
\includegraphics[width=0.5\columnwidth]{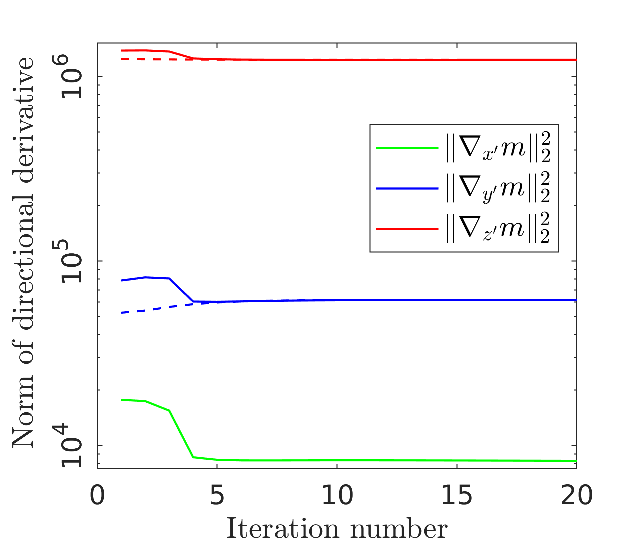}
\caption{3D Parihaka real data. Solid lines: evolution of the norm of directional derivatives in generating Figure \ref{Parihaka_angles} via Algorithm \ref{Algorithm3D}.
Dashed lines: the norm of derivatives when fully minimizing $\|\nabla_{\!\bold{x}'(\strike,\tilt)}\bold{m}\|_2^2$ at each iteration before updating the dip angle.}
\label{Parihaka_angles_ndrives}
\end{figure}

In our assessment of the proposed algorithm's performance, 
we use the Parihaka full-angle-stack seismic volume from New Zealand published by New Zealand Petroleum and Minerals. This dataset is freely accessible (\url{https://wiki.seg.org/wiki/Parihaka-3D}).
We select a segment of the data characterized by dipping layers (Figure \ref{Parihaka_angles}a). The chosen segment has dimensions of 201 (inline traces) $\times$ 201 (xline traces) $\times$ 101 (time samples). We initially employ Algorithm \ref{Algorithm3D} to estimate the orientation parameters; the  strike, tilt, and dip angles obtained after 20 iterations are displayed in Figures \ref{Parihaka_angles}b-\ref{Parihaka_angles}d. 
Observing the time slice of the data (Figure \ref{Parihaka_angles}a), we note that the layers are predominantly aligned along the (inline) $x$ axis, a trend accurately captured in the strike field, displaying small angles between $-8^\circ$ and $0^\circ$ (Figure \ref{Parihaka_angles}b). 
Additionally, we observe that the layers are nearly horizontal in the time-inline section with a small positive tilt that increases with depth. This information is correctly reflected in the tilt field (Figure \ref{Parihaka_angles}c), where the determined angles range between $0^\circ$ and $+8^\circ$. As the layers exhibit more dip along the xline, as evident from the visible time-xline section of the data, we expect more dip angles than strike and tilt. This expectation is accurately represented in the dip field (Figure \ref{Parihaka_angles}d), where the determined angles range between $+5^\circ$ and $+15^\circ$.
Figure \ref{Parihaka_angles_ndrives} shows the variation in the norm of directional derivatives. It is evident that the algorithm rapidly converges to a stationary point.

\begin{figure}[htb!] 
\begin{center}
\begin{tabular}{ccc}
\hspace{-0cm}\small{\textbf{a)}} & \hspace{-0cm}\small{\textbf{b)}} \\
\hspace{-0cm}\includegraphics[scale=0.4]{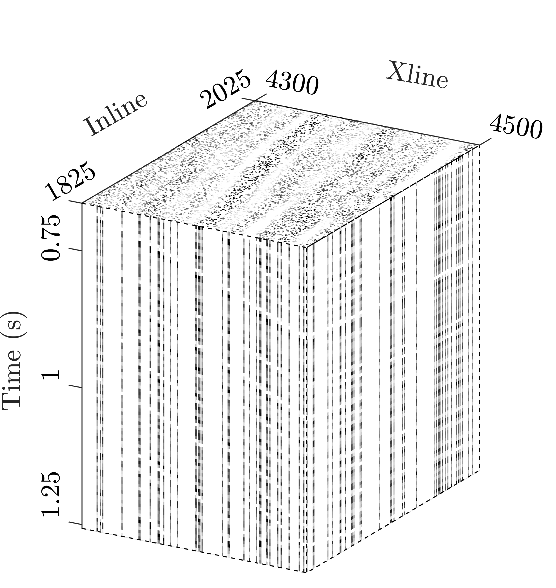} &
\hspace{-0cm}\includegraphics[scale=0.4]{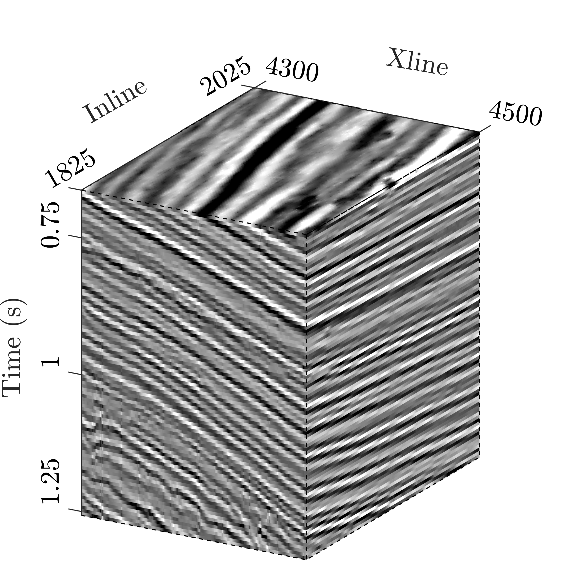} & \\
\hspace{-0cm}\small{\textbf{c)}}  &  \hspace{-0cm}\small{\textbf{d)}}\\
\hspace{-0cm}\includegraphics[scale=0.4]{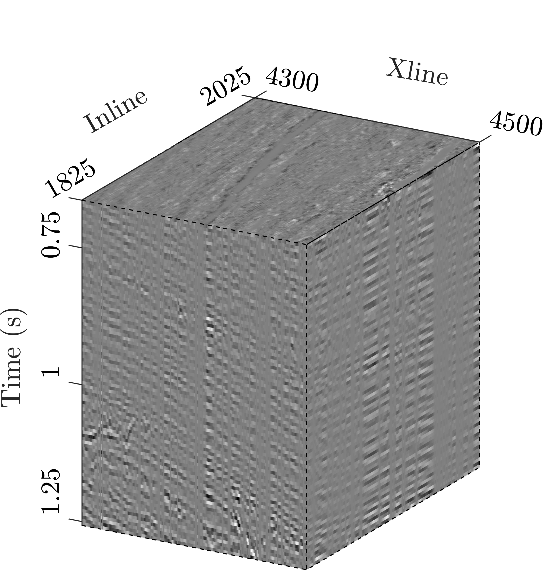} &
\hspace{-0cm}\includegraphics[scale=0.4]{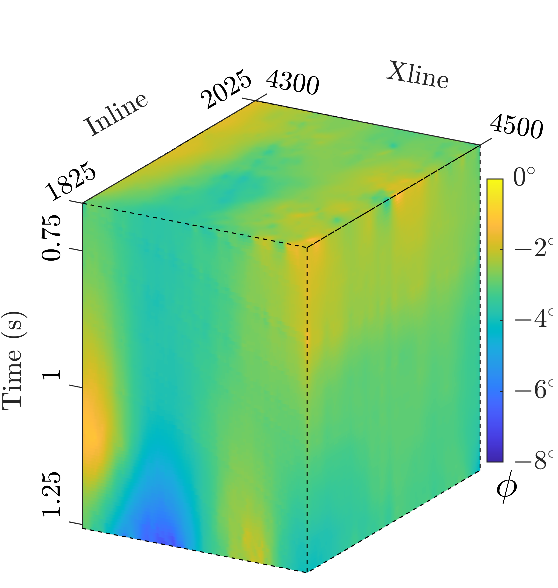} \\
\hspace{-0cm}\small{\textbf{e)}} & \hspace{-0cm}\small{\textbf{f)}}  \\
\hspace{-0cm}\includegraphics[scale=0.4]{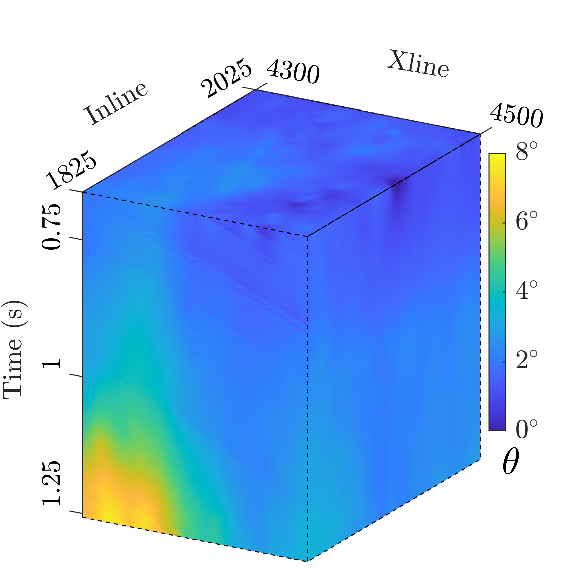} &
\hspace{-0cm}\includegraphics[scale=0.4]{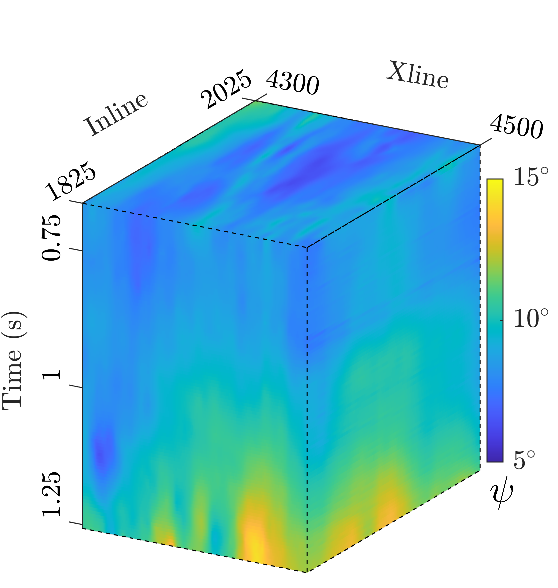}
\end{tabular}  
\caption{3D field data interpolation example. (a) 3D Parihaka field data with 80\% of the randomly chosen traces removed. (b) result of missing data interpolation by the 3D anisotropic Tikhonov regularization. (c) The difference between the reconstruction and the original data. (d-f) Estimated orientation parameters (d) strike angle $\bolds{\strike}$, (e) tilt angle $\bolds{\tilt}$, and (f) dip angle $\bolds{\dip}$.}
\label{Parihaka_interpolation}
\end{center}
\end{figure}

\begin{figure}
\center
\includegraphics[width=0.5\columnwidth]{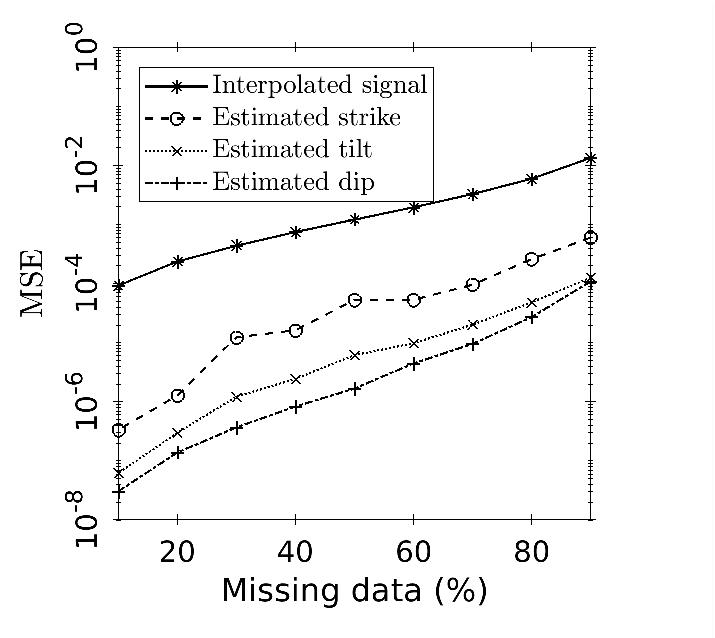}
\caption{Performance evaluation of Algorithm \ref{Algorithm3D} for simultaneous interpolation and estimation orientation parameters strike, tilt, and  dip for different percents of missing data. The MSE is calculated using the original signal and its associated orientation parameters as references for comparison (Figure \ref{Parihaka_angles}).}
\label{Parihaka_Interp_MSE}
\end{figure}

Finally, we harness local anisotropic regularization for a 3D missing trace interpolation experiment. This involved the random removal of 80\% of the traces from the original Parihaka data volume, generating the missing data volume displayed in Figure \ref{Parihaka_interpolation}a. The resulting interpolated data volume is illustrated in Figure \ref{Parihaka_interpolation}b, while the associated error relative to the original data is shown in Figure \ref{Parihaka_interpolation}c. Figures \ref{Parihaka_interpolation}d-\ref{Parihaka_interpolation}f show the strike, tilt, and dip angles estimated throughout the interpolation process.
To gain insights into the algorithm's performance under varying percentages of missing data, we applied the interpolation process with different levels of missing data. Figure \ref{Parihaka_Interp_MSE} shows the MSE for both the interpolated data and the estimated orientation parameters at different missing data levels, ranging from 10\% to 90\%. The original full data and its associated angles from Figure \ref{Parihaka_angles} serve as reference benchmarks. The consistent trend of the error curves in Figure \ref{Parihaka_Interp_MSE} indicates the algorithm stability in interpolation across different degrees of missing data.

%%%%%%%%%%%%%%%%%%%%%%%%%%%%%%%%%%%%%%%%%%%%%%%%%%%%%%%%%%%%%%%%%%%%%%%%%%%%%%%%%%%%%%%%%%%%%%%%

%%%%%
%%%%%%%%%%%%%%%%%%%%%%%%%%%%%%%%%%%%%%%%%%%%%%%%%%%%%%%%%%%%%%%%%%%%%%%%%%%%%%%%%%%%%%%%%%%%%%%%

\section{Discussion}
\noindent The introduction of local anisotropic regularization by \citet{Li_2000_IGD} and \citet{Clapp_2004_IGI} marked a significant advancement in geophysical inverse problems, allowing for the incorporation of geological information to ensure a unique and meaningful solution.  Previous approaches operated under the assumption of a priori knowledge of orientation information, and often estimated such information separately before inversion using methods like filter techniques based on plane wave destruction. However, such approaches may encounter limitations when dealing with under-determined problems, where it is difficult to get a good estimate of the solution beforehand. 

In this study, we relaxed this constraint by proposing a method that dynamically estimates orientation angles during the inversion process. This innovation offers two distinct advantages: firstly, anisotropic regularization becomes applicable to a wide range of problems, including under- and over-determined cases, and in any domain (time or depth). The required orientation parameters for designing the regularization function evolve adaptively with model estimation. Secondly, the recovered angles serve as a byproduct of the inversion, contributing valuable information for interpretation purposes.

Regularization primarily aims to suppress fluctuating, noise-like features in the estimated model that may arise from noise in the input data and the problem's ill-conditioning. Specifically, Tikhonov regularization in \eref{eq:iso} incorporates prior assumptions about the desired solution's smoothness \citep{Hansen_1998_RDD}. This is achieved using homogeneous smoothing kernels, where each point in the model is computed with a weighted average of neighboring points, with weights decreasing as the distance from the target point increases.
However, this traditional approach to regularization is somewhat naive as it does not consider the local structure of the model. In contrast, our proposed local anisotropic regularization adapts the shape of the smoothing kernel to apply smoothing along the structures present in the model. Namely, the adaptive smoothing kernels align with variations in the orientation and coherence paths of model features. Because signal variations are minimal along these coherence paths, the signal is preserved while noise is effectively canceled out through the averaging process of the kernel.

Numerous synthetic and field examples demonstrate that the new anisotropic regularization method provides solutions to geophysical inverse problems with high resolution and accurate representations of true structures. However, these advantages are accompanied by increased computational costs compared to standard isotropic regularization. Indeed, the latter just solves a problem expressed in the 2-norm (see equation \ref{eq:iso}, for which a direct solution method can be applied (although an iterative solver has to be adopted for large-scale problems).
The new alternating optimization framework for the model and orientation angles necessitates solving the original problem multiple times to refine the orientation parameter estimates gradually. However, this repeated solution of the model subproblem is also necessary in standard isotropic regularization to determine an appropriate value for the fitting parameter $\mu$, following a continuation strategy \citep{Hansen_1998_RDD}. Thus, the primary computational cost of our new algorithm for anisotropic regularization is the computation of the Tikhonov filter applied at each iteration (subproblem \ref{sub_theta} for 2D problems and subproblems \ref{sub_strike3d}-\ref{sub_dip3d} for 3D problems) to update the orientation parameters. Fortunately, the diagonal structure of the Jacobian matrix for these filters allows for efficient computation using standard linear algebra tools.  The additional cost of the new approach compared to standard isotropic regularization is justified because, in addition to achieving higher quality reconstructions in practical applications, the regularization parameters are adaptively estimated.
%(indeed, determining the optimum regularization parameter for the standard isotropic regularization may involve solving the regularized problem for a range of parameter values, following a continuation strategy, leading to an increased computational cost).
% to find a suitable value
% %Despite this additional cost, it is justified in practical applications where the regularization parameter is unknown. 
% Even for the standard isotropic regularization, . 
% The proposed algorithm updates orientation angles and the regularization parameter simultaneously, simplifying the computational process. 

The regularization parameter $\rho$ plays a critical role in controlling the stationarity of the orientation angles, designing the regularization function. Proper selection of its value is crucial for the algorithm's convergence to a reasonable solution, particularly when dealing with underdetermined problems like trace interpolation. A recommended strategy is to initiate with a relatively large value for $\rho$ or $\epsilon_1$ and subsequently decrease it gradually throughout the iterations. While numerical results indicate the effectiveness of a fixed value of $\rho$, additional research is warranted to develop a robust algorithm for the automatic selection of $\rho$. Potential strategies include a robust continuation approach or integration of learning algorithms with training data \citep{Chung_2024_ELM}. 

 The inherent $\ell_2$ nature of the proposed anisotropic regularization may introduce some blurring effects in discontinuities such as faults and reflection terminations. This limitation could be mitigated adopting a number of alternative strategies, which extend the approach proposed in this paper. First, and most straightforwardly, one may be exploring alternative norms, such as the $\ell_1$-norm, leading to a directional total variation regularizer, which also allows us to recover discontinuities along the directions where regularization is applied. Second, one may consider locally adjusting the values for the anisotropy parameters $\sigma_x$, $\sigma_y$ and $\sigma_z$, which automatically account for structures that may be more or less strongly aligned in the recovered directions. Third (generalizing the first and related to the second strategies), one may define a spatially varying norm regularization. For instance, employing an $\ell_p$ norm with spatially varying $p$ (within the range of [1, 2]) could be considered, as in  \citep{Han_2023_IRL}. Such adaptive norm selection enables adjustments to the contribution of directional derivatives in the regularization term, potentially providing a more effective means of imposing smoothness and preserving discontinuities in the subsurface structure.

Finally, in this paper, we used the least-squares method to measure the data misfit, assuming that the noise in the data follows a Gaussian distribution. However, in practice, the data may be contaminated by outliers or erratic noises. In such cases, a robust error measure, such as the vector 1-norm, can be employed to impose smaller penalties on outliers.

\section{Conclusions}
In this paper we proposed a new method that  allows us to simultaneously recover model and orientation parameters by applying local anisotropic Tikhonov regularization, where the latter are used to define an effective regularizer for the former. The method proved to be extremely valuable in a variety of test problems from geophysics. 
Future work will involve deriving strategies for selecting the regularization parameter $\rho$ for the orientation parameters and extending the current approach to non-smooth regularizers,  such as directional total variation (as explained in the previous section). Additionally, we aim to apply these methods to more challenging and potentially nonlinear problems, such as waveform inversion. %{Depending on what we include about FWI: mention a more consistent treatment of nonlinear inverse problems. Are there other interesting test problems in geophysics not considered here?}

%\section{ACKNOWLEDGMENTS}  
%This research was financially support by the SONATA BIS grant (No. 2022/46/E/ST10/00266) of the National Science Center in Poland. 

\section{Appendix A: Estimation of the Orientation Parameter for a Plane Wave} \label{appA}
Consider a 2D sinusoidal plane wave $m(x,z)= \cos(k_xx-k_zz)$ with $k_x=k\sin\tilt^*$, $k_z=k\cos\tilt^*$, $k=\sqrt{k_x^2+k_z^2}$, and $\tilt^*\in (-\frac{\pi}{2},\frac{\pi}{2}]$ (see Figure \ref{slope2}). The gradient of $m(x,z)$ is
\begin{equation}
\begin{pmatrix}
\partial_{\!x}m\\
\partial_{\!z}m
\end{pmatrix}=
\begin{pmatrix}
-k_x\sin(k_xx-k_zz)\\
+k_z\sin(k_xx-k_zz)
\end{pmatrix}=
\begin{pmatrix}
-c\sin\tilt^*\\
c\cos\tilt^*
\end{pmatrix},
\end{equation}
where $c=k\sin(k_xx-k_zz)$. Without loss of generality, we assume $c=1$.
Consider the weighted norm 
\begin{equation}
p(\tilt)=\|\nabla_{\!\tilt}m(x,z)\|_{\Sigma}^2
\end{equation}
of the directional derivative of $m(x,z)$ with respect to $\theta$. 
 By taking $\sigma_x=1$ and $\sigma_z=0$ we have 
\begin{equation}
p(\tilt)=(-\sin\tilt^*\cos\tilt + \cos\tilt^*\sin\tilt)^2,
\end{equation} 
whose first and second derivatives are
\begin{equation}
\frac{dp}{d\tilt}=-\sin(2\tilt^* - 2\tilt),\quad \frac{d^2p}{d\tilt^2}=2\cos(2\tilt^* - 2\tilt).
\end{equation}
The first derivative vanishes at $\tilt = \tilt^* \pm n\frac{\pi}{2}$ for $n\in\mathbb{Z}$ (see again Figure \ref{slope2}). Furthermore, $\frac{d^2p}{d\tilt^2}\geq 0$ for $\tilt \in (\tilt^* -\frac{\pi}{2},\tilt^* +\frac{\pi}{2})$, thus $p(\tilt)$ is locally convex and gradient-based algorithms starting from $0 \in (\tilt^* -\frac{\pi}{2},\tilt^* +\frac{\pi}{2})$ can converge to the true solution.

\section{Appendix B: Tilt Estimation in 2D by ADMM} \label{appB}
Here we provide the procedure for addressing the tilt estimation problem (as defined by \eref{main_theta}) using ADMM.
By introducing the auxiliary variable $\bold{z}$, the constraint $\bolds{z}=\bolds{\tilt}$, and the Lagrange multiplier vector $\bolds{\lambda}$, the ADMM formulation results in the following iterative scheme for estimating the tilt field $\bolds{\tilt}$ associated with a given model $\bold{m}$ \citep[see][ for more details]{Tapia_1977_DMM,Gabay_1976_ADA,Boyd_2011_DOS} 
\begin{subequations} \label{theta_app}
\begin{align} 
\bolds{\tilt}^{+}&=\arg\min_{\bolds{\tilt}} ~\frac12\sum_{i=1}^N\| \bold{R}(\bolds{\tilt}_i)(\tilde{\nabla}\bold{m})_i\|_{\bold{\Sigma}}^2+ \frac{\rho}{2}\|\nabla\bolds{\tilt}\|_2^2\nonumber \label{sub_theta_app} \\
& \hspace{1.75cm}-\langle\bolds{\lambda},\bolds{\tilt}-\bold{z}\rangle+ \frac{\tau}{2}\|\bolds{\tilt}-\bold{z}\|_2^2,\\
\bold{z}^{+} &= \min(\max(\bolds{\tilt}^+-\bolds{\lambda},-\frac{\pi}{2}),\frac{\pi}{2}), \\
\bolds{\lambda}^{+}& =\bolds{\lambda} - \tau(\bolds{\tilt}^{+}-\bold{z}^{+}).
\end{align}
\end{subequations}
where the triplet $(\bolds{\tilt},\bold{z},\bolds{\lambda})$ is the values at the current iteration, $(\bolds{\tilt}^{+},\bold{z}^{+},\bolds{\lambda}^{+})$ are the updated values, and $\tau>0$ is the penalty parameter. 

We address the problem in \eqref{sub_theta_app} through a single iteration of the Gauss-Newton method. 
Noting that
\begin{equation}
    \sum_{i=1}^N\| \bold{R}(\bolds{\tilt}_i)(\tilde{\nabla}\bold{m})_i\|_{\bold{\Sigma}}^2 = \|\tilde{\bolds{D}}(\bolds{\tilt})\bold{m}\|_2^2,
\end{equation}
where $\tilde{\bolds{D}}$ is defined as in \eqref{D_aniso}, but computed with the smooth derivative operators $\tilde{\nabla}_{\! x}$ and $\tilde{\nabla}_{\! z}$, and
assuming that the smooth derivative term $\tilde{\bolds{D}}(\bolds{\tilt})\bold{m}$ is differentiable at $\tilt$, we can approximate it using the first two terms of a Taylor series expansion:
\begin{equation}
\tilde{\bolds{D}}(\underbrace{\bolds{\tilt}+\delta\bolds{\tilt}}_{\bolds{\tilt}^+})\bold{m} \approx \tilde{\bolds{D}}(\bolds{\tilt})\bold{m} + \underbrace{\frac{\partial (\tilde{\bolds{D}}(\bolds{\tilt})\bold{m})}{\partial \tilt }}_{\bold{J}(\bold{m},\bolds{\tilt})}\delta\tilt.
\end{equation}
This gives the following closed form solution for the updated tilt field:
\begin{equation} \label{theta_sub_app}
\bolds{\tilt}^{+} = (\bold{J}(\bold{m},\bolds{\tilt})^T\!\bold{J}(\bold{m},\bolds{\tilt})+\tau I+\rho \nabla^T\nabla)^{-1}\!
\bold{g}(\bold{m},\bolds{\tilt},\bold{z},\bolds{\lambda}), 
\end{equation}
where $\bold{J}$, the Jacobian matrix, and $\bold{g}$ are defined as:
\begin{align} 
\bold{J}(\bold{m},\bolds{\tilt}) = 
\begin{pmatrix}
   \text{diag}(
\{\langle(\tilde{\nabla} \bold{m})_i,\frac{\partial \bold{x}'(\bolds{\tilt}_i)}{\partial\bold{\tilt}_i}\rangle\}_{i=1}^N)\\
    \text{diag}(\{\langle\sqrt{\epsilon}(\tilde{\nabla} \bold{m})_i,\frac{\partial \bold{z}'(\bolds{\tilt}_i)}{\partial\bold{\tilt}_i}\rangle\}_{i=1}^N)
\end{pmatrix},
\end{align}
and
\begin{equation}  \label{bk_app}
\bold{g}(\bold{m},\bolds{\tilt},\bold{z},\bolds{\lambda})= \bold{J}(\bold{m},\bolds{\tilt})^T[\bold{J}(\bold{m},\bolds{\tilt}) \bolds{\tilt} - \tilde{\bolds{D}}(\bolds{\tilt})\bold{m}]+
\tau \bold{z}+\bolds{\lambda}.
\end{equation}

\section{Appendix C: Dip, Strike, and Tilt Estimation in 3D by ADMM} \label{appC}
Similar to the 2D case discussed above, after introducing the auxiliary variables $\bold{z}_\bolds{\strike}$, $\bold{z}_\bolds{\tilt}$, $\bold{z}_\bolds{\dip}$ and the constraints $\bold{z}_\bolds{\strike}=\bolds{\strike}$, $\bold{z}_\bolds{\tilt}=\bolds{\tilt}$, $\bold{z}_\bolds{\dip}=\bolds{\dip}$, we employ ADMM to solve \eqref{main3d_theta} sequentially, with each angle updated independently at each iteration, as follows:
%{
\begin{subequations}
\begin{argmini}
      {\bolds{\strike}}
{\hspace{-0.5cm} \frac12\sum_{i=1}^N\| \bold{R}(\bolds{\strike}_i,\bolds{\tilt}_i,\bolds{\dip}_i)(\tilde{\nabla} \bold{m}^+)_i\|_{\bolds{\Sigma}}^2}
      {\label{main3d_strike}}{\bolds{\strike}^+=}
       \breakObjective{\hspace{-0.5cm} + \frac{\rho_{\strike}}{2}\|\nabla\bolds{\strike}\|_2^2-\langle\bolds{\lambda}_{\strike},\bolds{\strike}-\bold{z}_{\strike}\rangle+ \frac{\tau_\strike}{2}\|\bolds{\strike}-\bold{z}_{\strike}\|_2^2}
\end{argmini}
\begin{argmini}
      {\bolds{\tilt}}
{\hspace{-0.5cm} \frac12\sum_{i=1}^N\| \bold{R}(\bolds{\strike}_i^+,\bolds{\tilt}_i,\bolds{\dip}_i)(\tilde{\nabla} \bold{m}^+)_i\|_{\bolds{\Sigma}}^2}
      {\label{main3d_tilt}}{\bolds{\tilt}^+=}
       \breakObjective{\hspace{-0.5cm} + \frac{\rho_{\tilt}}{2}\|\nabla\bolds{\tilt}\|_2^2-\langle\bolds{\lambda}_{\tilt},\bolds{\tilt}-\bold{z}_{\tilt}\rangle+ \frac{\tau_\tilt}{2}\|\bolds{\tilt}-\bold{z}_{\tilt}\|_2^2}
\end{argmini}
\begin{argmini}
      {\bolds{\dip}}
{\hspace{-0.5cm} \frac12\sum_{i=1}^N\| \bold{R}(\bolds{\strike}_i^+,\bolds{\tilt}_i^+,\bolds{\dip}_i)(\tilde{\nabla} \bold{m}^+)_i\|_{\bolds{\Sigma}}^2}
      {\label{main3d_dip}}{\bolds{\dip}^+=}
       \breakObjective{\hspace{-0.5cm} + \frac{\rho_{\dip}}{2}\|\nabla\bolds{\dip}\|_2^2-\langle\bolds{\lambda}_{\dip},\bolds{\dip}-\bold{z}_{\dip}\rangle+ \frac{\tau_\dip}{2}\|\bolds{\dip}-\bold{z}_{\dip}\|_2^2}
\end{argmini}
\begin{align}
\bold{z}_{\strike}^{+} &= \min(\max(\bolds{\strike}^{+}-\bolds{\lambda}_{\strike},-\frac{\pi}{2}),\frac{\pi}{2}),\\
\bold{z}_{\tilt}^{+} &= \min(\max(\bolds{\tilt}^{+}-\bolds{\lambda}_{\tilt},-\frac{\pi}{2}),\frac{\pi}{2}),\\
\bold{z}_{\dip}^{+} &= \min(\max(\bolds{\dip}^{+}-\bolds{\lambda}_{\dip},-\frac{\pi}{2}),\frac{\pi}{2}), \\
\bolds{\lambda}_{\strike}^{+}    &=\bolds{\lambda}_{\strike} - \tau_{\strike} (\bolds{\strike}^{+}-\bold{z}_{\strike}^{+}),\\
\bolds{\lambda}_{\tilt}^{+}    &=\bolds{\lambda}_{\tilt} - \tau_{\tilt} (\bolds{\tilt}^{+}-\bold{z}_{\tilt}^{+}),\\
\bolds{\lambda}_{\dip}^{+}  &=\bolds{\lambda}_{\dip} - \tau_{\dip} (\bolds{\dip}^{+}-\bold{z}_{\dip}^{+}),
\end{align}
\end{subequations}
%}
where $\bullet$ are the values at the current iteration, $\bullet^{+}$ are the updated values, and $\tau_{\strike}$, $\tau_{\tilt}$, $\tau_{\dip}$ are penalty parameters. 

Now, we explain how to solve subproblems \ref{main3d_strike}-\ref{main3d_dip}.
As for the 2D problems, we solve subproblems \ref{main3d_strike}-\ref{main3d_dip} through a single iteration of the Gauss-Newton method.
Noting that
\begin{equation}
 \sum_{i=1}^N\| \bold{R}(\bolds{\strike}_i,\bolds{\tilt}_i,\bolds{\dip}_i)(\tilde{\nabla} \bold{m})_i\|_{\bolds{\Sigma}}^2 = \|\tilde{\bolds{D}}(\bolds{\strike},\bolds{\tilt},\bolds{\dip})\bold{m}\|_2^2,
\end{equation}
where $\tilde{\bolds{D}}$ is defined as \eqref{D3D} but computed with the smooth derivative operator $\tilde{\nabla}$, and
assuming that the smooth derivative term $\tilde{\bolds{D}}(\bolds{\strike},\bolds{\tilt},\bolds{\dip})\bold{m}$ is differentiable at $\bolds{\strike},\bolds{\tilt},\bolds{\dip}$, we can approximate it using the first two terms of a Taylor series expansion:
\begin{equation*}
\begin{cases}
\tilde{\bolds{D}}(\bolds{\strike}+\delta\bolds{\strike},\bolds{\tilt},\bolds{\dip})\bold{m} \approx \tilde{\bolds{D}}(\bolds{\strike},\bolds{\tilt},\bolds{\dip})\bold{m} + \frac{\partial (\tilde{\bolds{D}}(\bolds{\strike},\bolds{\tilt},\bolds{\dip})\bold{m})}{\partial \bolds{\strike}}\delta\bolds{\strike},\\
\tilde{\bolds{D}}(\bolds{\strike},\bolds{\tilt}+\delta\bolds{\tilt},\bolds{\dip})\bold{m} \approx \tilde{\bolds{D}}(\bolds{\strike},\bolds{\tilt},\bolds{\dip})\bold{m} + \frac{\partial (\tilde{\bolds{D}}(\bolds{\strike},\bolds{\tilt},\bolds{\dip})\bold{m})}{\partial \bolds{\tilt}}\delta\bolds{\tilt},\\
\tilde{\bolds{D}}(\bolds{\strike},\bolds{\tilt},\bolds{\dip}+\delta\bolds{\dip})\bold{m} \approx \tilde{\bolds{D}}(\bolds{\strike},\bolds{\tilt},\bolds{\dip})\bold{m} + \frac{\partial (\tilde{\bolds{D}}(\bolds{\strike},\bolds{\tilt},\bolds{\dip})\bold{m})}{\partial \bolds{\dip}}\delta\bolds{\dip}.
\end{cases}
\end{equation*}
This allows us to express the updated angles $\bolds{\strike}^{+}=\bolds{\strike} + \delta \bolds{\strike}$, $\bolds{\tilt}^{+}=\bolds{\tilt}+\delta\bolds{\tilt}$ and $\bolds{\dip}^{+}=\bolds{\dip}+\delta\bolds{\dip}$ in the following closed forms:
\begin{subequations}\label{strike_sub}
\begin{align} 
\bolds{\strike}^{+} &=\! \bold{H}_{\strike}(\bolds{m}^+,\bolds{\strike},\bolds{\tilt},\bold{\dip})^{-1}\bold{g}_{\strike}(\bolds{m}^+,\bolds{\strike},\bolds{\tilt},\bold{\dip},\bold{z}_{\strike},\bolds{\lambda}_{\strike}), \\
\bolds{\tilt}^{+} &=\! \bold{H}_{\tilt}(\bolds{m}^+,\bolds{\strike}^+,\bolds{\tilt},\bold{\dip})^{-1}\bold{g}_{\tilt}(\bolds{m}^+,\bolds{\strike}^+,\bolds{\tilt},\bold{\dip},\bold{z}_{\tilt},\bolds{\lambda}_{\tilt}), \\
\bolds{\dip}^{+} &=\! \bold{H}_{\dip}(\bolds{m}^+\!,\bolds{\strike}^+\!,\bolds{\tilt}^+\!,\bold{\dip})^{-1}\bold{g}_{\dip}(\bolds{m}^+\!,\bolds{\strike}^+\!,\bolds{\tilt}^+\!,\bold{\dip},\bold{z}_{\dip},\bolds{\lambda}_{\dip}).
\end{align}
\end{subequations}
In these equations, the Hessian matrices $\bold{H}_{\bullet}$ and right-hand-side vectors $\bold{g}_{\bullet}$ are defined as
\begin{align} \label{Hess3d}
\bold{H}_{\bullet}(\bolds{m},\bolds{\strike},\bolds{\tilt},\bold{\dip})=&\bold{J}_{\bullet}(\bolds{m},\bolds{\strike},\bolds{\tilt},\bold{\dip})^T\bold{J}_{\bullet}(\bolds{m},\bolds{\strike},\bolds{\tilt},\bold{\dip})
+\tau_{\bullet}I+\rho_{\bullet} \nabla^T\nabla
\end{align}
and
\begin{align}  \label{g_strike}
\bold{g}_{\bullet}(\bolds{m},\bolds{\strike},\bolds{\tilt},\bolds{\dip},&\bold{z},\bolds{\lambda})=
\bold{J}_{\bullet}(\bolds{m},\bolds{\strike},\bolds{\tilt},\bolds{\dip})^T
[\bold{J}_{\bullet}(\bolds{m},\bolds{\strike},\bolds{\tilt},\bolds{\dip})\bold{\bullet}
-\tilde{\bold{D}}(\bolds{\strike},\bolds{\tilt},\bolds{\dip})\bold{m}]
+\tau_{\bullet}\bold{z}+\bolds{\lambda},
\end{align}
respectively, where the Jacobian matrices $\bold{J}_{\bullet}$ are defined as:
\begin{align} \label{J}
\bold{J}_{\bullet}(\bolds{m},\bolds{\strike},\bolds{\tilt},\bolds{\dip})=
\begin{pmatrix}
  \text{diag}(\langle(\tilde{\nabla} \bold{m})_i,\frac{\partial \bold{x}'(\bolds{\strike}_i,\bolds{\tilt}_i,\bold{\dip}_i)}{\partial\bold{\bullet}_i}\rangle)  \\
   \text{diag}(\langle \sqrt{\delta}(\tilde{\nabla} \bold{m})_i,\frac{\partial \bold{y}'(\bolds{\strike}_i,\bolds{\tilt}_i,\bold{\dip}_i)}{\partial\bold{\bullet}_i}\rangle) \\
    \text{diag}(\langle \sqrt{\epsilon}(\tilde{\nabla} \bold{m})_i,\frac{\partial \bold{z}'(\bolds{\strike}_i,\bolds{\tilt}_i,\bold{\dip}_i)}{\partial\bold{\bullet}_i}\rangle)  
\end{pmatrix}.
\end{align}
Here $\bullet$ stands for $\strike$, $\tilt$ and $\dip$. 
% The proposed algorithm for 3D problems is summarized in Algorithm \ref{Algorithm3D}. Note that, to follow the procedure just described, one should take $K=1$. 

\section{ACKNOWLEDGMENTS}  
This research was partially funded by the SONATA BIS grant (No. 2022/46/E/ST10/00266) of the National Science Center in Poland. 

\newcommand{\SortNoop}[1]{}

\end{document}